# LINE: Public-key encryption

Gennady Khalimov[1], Yevgen Kotukh[2],

1 Kharkiv National University of Radioelectronics, Kharkiv, Ukraine
gennady.khalimov@gmail.com

2 Dnipro University of Technology, Dnipro, Ukraine
yevgenkotukh@gmail.com

**Abstract.** We propose a public key encryption cryptosystem based on solutions of linear equation systems with predefinition of input parameters through shared secret computation for factorizable substitutions. The existence of multiple equivalent solutions for an underdetermined system of linear equations determines the impossibility of its resolution by a cryptanalyst in polynomial time. The completion of input parameters of the equation system is implemented through secret homomorphic matrix transformation for substitutions factorized over the basis of a vector space of dimension m over the field $F_2$. Encryption is implemented through computation of substitutions that are one-way functions on an elementary abelian 2-group of order $2^m$. Decryption is implemented through completion of input parameters of the equation system. Homomorphic transformations are constructed based on matrix computations. Matrix computations enable the implementation of high security and low computational overhead for homomorphic transformations.

## Introduction

The main task formulated in the NIST project is the standardization of KEMs and signatures with low overhead for keys, signatures, and computation time [1]. Based on the results of the NIST PQC standardization project, the best results in the key encapsulation category are demonstrated by the algorithms: CRYSTALS-Kyber [2], Classic McEliece [3], and HQC [4], and in the digital signature category: Crystals-Dilithium (Dilithium) [5], Falcon [6-8], and SPHINCS+. The design principles and security problems underlying these algorithms are derived from lattice-based cryptography, error-correcting code theory, and hash-based schemes.

The security of lattice-based cryptography is achieved through the use of NP-hard problems such as finding shortest vectors (SVP, CVP, SVIP) and learning with errors (LWE, LWR) [9-12]. To ensure security, Dilithium relies on the Fiat-Shamir structure and Aborts, as well as SVP [13]. SPHINCS+ relies exclusively on assumptions about the hardness of hash functions. These assumptions are perceived as much more conservative than the structured assumptions underlying Dilithium and Falcon. Overall, the NIST-selected PQC candidates Kyber and Dilithium are considered secure and efficient schemes.

Computational cost and parameter size estimates for post-quantum KEM schemes are provided in NIST report [1]. The security of Kyber has been thoroughly analyzed and is based on a solid foundation of lattice-based cryptography results. Kyber has excellent overall performance with respect to software, hardware, and many hybrid settings. For implementation costs of 256-bit cryptography, Kyber requires public keys of 1568 bytes, secret keys of 3168 bytes, ciphertext of 1568 bytes, encryption costs of 97,000 cycles, and decryption costs of 80,000 cycles. Dilithium requires public keys of 2600 bytes for implementation, generates signatures of 4600 bytes, signing costs of 345,000 cycles, and signature verification of 150,000 cycles. SPHINCS+ has much worse performance than other standards: for example, signature size, verification time, and signing time are respectively one, two, and three orders of magnitude higher than, say, Dilithium. Classic McEliece requires the highest computational costs and the highest communication cost due to large public key size while having the smallest ciphertext. Classic McEliece is the slowest scheme for key generation, and HQC is the slowest for encapsulation and decapsulation. The fastest scheme is Kyber.

Large overhead costs are determined by the fact that solving the problem of cryptographic secrecy requires significant expansion of the ciphertext space compared to the plaintext space. For cryptosystems based on NP-hard problems, this is an inevitable solution that leads to an actual increase in operational costs compared to AES256 encryption by tens of times (49 times). Cryptosystems of this type do not have provable security against quantum cryptanalysis, and it can be assumed that this will be a persistent threat. PQC schemes that do not exploit the complexity problem in direct formulation have other constructive solutions. Thus, SPHINCS+ is built on assumptions about the hardness of hash functions and exploits the idea of one-time secret pads. After using a secret (input value for which a hash code was computed), the next secret is used, and so on. The Classic McEliece cryptosystem is built on matrix computations structured by a generator matrix of an error-correcting redundant code. Attacks are reduced to solving a brute-force problem of decoding the ciphertext. The price for quantum secrecy is large overhead for common parameters and cryptosystem keys for large ciphertext, as in the case of SPHINCS+, as well as large operational costs for storage, transmission over channels, and computation time for Classic McEliece.

To solve the problem of constructing a post-quantum cryptosystem with low implementation costs and satisfying NIST security requirements, we propose building public-key cryptosystems with a new concept based on brute-force problems with equiprobable solutions for incomplete systems of linear equations and applying secret sharing over ciphertexts for completion of these equations. Secret sharing is one of the cryptographic mechanisms. An example is Shamir's threshold scheme based on polynomial approximation by its values. The secrecy of Shamir's scheme is guaranteed by the properties of polynomial algebra, and an attack on the common key is only brute-force. The condition when the number of equations is less than the number of input parameters leads to an incomplete system of linear equations with respect to unknowns (input text) and the impossibility of its resolution by a cryptanalyst in polynomial time.

**Our Contributions**

We develop the theory of constructing asymmetric cryptosystems with secrecy that is determined by the conditions of brute-force problems. As the foundation for constructing such a cryptosystem, we adopted the property of an incomplete system of linear equations with respect to its solutions. Since a unique solution exists only for a fully determined system of linear equations, we defined a mechanism for parametric completion of the equation system through secret homomorphic transformations of ciphertexts. We developed the theory of secret sharing over ciphertexts based on homomorphic matrix transformations over factorized substitutions. We applied factorized substitutions that act as secret one-way substitutions. The one-way property of substitutions is characterized by direct keyless transformation and secret inverse transformation, which is a necessary condition for constructing public key encryption. The potential secrecy of a cryptosystem based on an incomplete system of linear equations is determined by the cardinality of the solution set of the equation system, and an attack on the ciphertext is only brute-force.

**Organization**

In the next section, we present a description of the LINE cryptosystem based on matrix computations, secret sharing, and key substitutions for plaintext. In the third section, we present secret one-way substitutions on an elementary abelian 2-group of order $2^m$. After, we describe secret sharing in the LINE cryptosystem based on homomorphic transformation with the property that the action of the inverse transformation for any input vector leads to a key vector. Next, we describe the LINE scheme for public key encryption in a cryptosystem with linear equations. In the last section, we performed security analysis, complexity estimates of main brute-force attacks and analytical attacks. In the Appendix, we provide an example of public key encryption computation in the LINE cryptosystem.

**LINE: a cryptosystem based on an incomplete system of linear equations**

To construct the LINE cryptosystem, we utilize the well-known fact that an underdetermined system of linear equations has multiple solutions. Let the system of linear equations be described by a binary matrix $A[l \times k]$, $l < k$, which connects the values of the input vector $y[k]$ with the output vector $u[l]$

$$A \times y = u, \qquad (1)$$

where are vectors $y$ and $u$ have dimensions accordingly $k$ and $l$ with $m$ bit components. The calculations in equation (1) are performed using the bitwise XOR operation on $m$ bit components of the vector $y$.

The solution of equation (1) has a maximum uncertainty relatively $y[k] \in Y$ equal to $|Y| = 2^{(k-l)m}$, $l < k \leq 2l$. Direct guessing of the solution has a probability of $2^{-(k-l)m}$. The application of an underdetermined system of linear equations for cryptosystem construction potentially provides high security and good operational characteristics through parameter selection $k$, $l$ and $m$. Let's write the vector $y[k]$ in the form

$$y[k] = y[l] \| y[l,k] = [y_1, y_2, \ldots y_l] \| [y_{l+1}, y_{l+2}, \ldots y_k].$$

The solution of equation (1) requires redetermining $k - l$ the components of the vector $y[l,k] = [y_{l+1}, y_{l+2}, \ldots y_k]$.

To build a cryptosystem, we define the following requirements for $y[l,k]$.

1. *Secrecy* $y[l,k]$. The uncertainty of the solution of equation (1) is determined by the uncertainty of the values of the components of the vector $y[l,k]$, and therefore $y[l,k]$ must be a secret key.

2. *Invariance* to the values of the input text. Let us represent $x$ as $k$ a component vector $x[k] = [x_1, x_2, \ldots x_k]$. Let the mapping $\beta$ be a vector-to-vector $y$ transformation $x$

$$\beta : x \rightarrow [\beta_1(x_1), \ldots, \beta_k(x_k)] = [y_1, \ldots, y_k]. \tag{2}$$

Invariance allows us to obtain a solution to equation (1) for the components $y[l]$ at different input vectors $x$.

***Solution for implementing requirements for*** $y[l,k]$.

The key secrecy requirement $y[l,k]$ can be satisfied based on a secret sharing scheme for the vector $y[k]$. Let the mapping $\sigma$ be a secret homomorphic transformation

$$\sigma : y \rightarrow y^*, \tag{3}$$

where $y^* = [y_1^*, \ldots, y_q^*]$ is the set of $m$ bit $k$ component vectors $y_i^* = [y_{i1}^*, y_{i2}^*, \ldots y_{ik}^*]$, $i = \overline{1,q}$.

Let's define vectors $y^* = [y_1^*, \ldots, y_q^*]$ as partial secrets. The mapping $\sigma^{-1} : y^* \rightarrow y$ is a conjugate homomorphic transformation. For a set of vectors $y^* = [y_1^*, \ldots, y_q^*]$ it is possible to calculate $q$ the cipher of texts $[u_1, \ldots, u_q]$ using equation (1)

$$u_1 = A \times y_1^*$$
$$u_2 = A \times y_2^*$$
$$\ldots\ldots\ldots\ldots$$
$$u_q = A \times y_q^* \tag{4}$$

To decrypt $[u_1,...,u_q]$, we take into account that, due to the linearity of equations (4), the action of the relative transformation $\sigma^{-1}$ For $y^* = [y_1^*,...,y_q^*]$ transferred to cipher texts $u_i$

$$u_\sigma = \sigma^{-1}(u_1,...,u_q). \tag{5}$$

As a result, we obtain an equation for $y$

$$A \times y = u_\sigma. \tag{6}$$

The solution of equation (6) is possible if the secret vector is known $y[l,k] = [y_{l+1}, y_{l+2},...y_k]$, which will give the desired value $y[l] = [y_1, y_2,...y_l]$. Vector of values $y[l] = [y_1, y_2,...y_l]$ is defined as a shared secret in a secret sharing scheme.

The block diagram for solving an incomplete system of linear equations with supplementation of input parameters through shared secret computation is presented in Fig. 1.

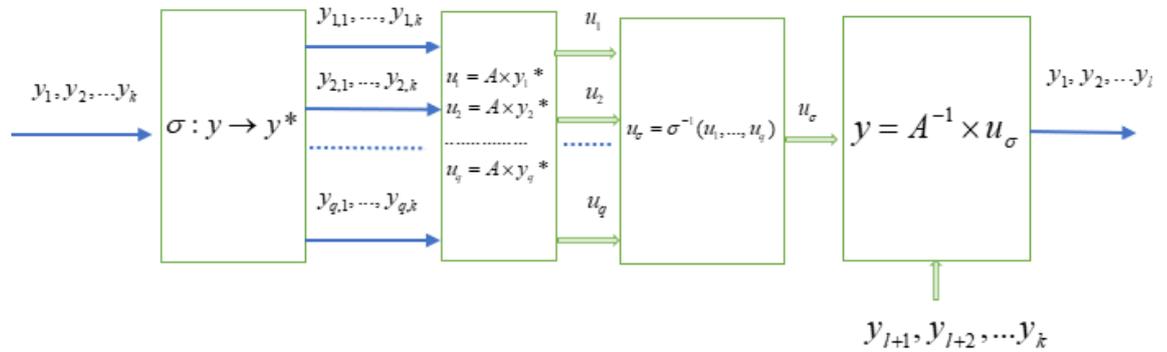

Fig. 1. Block diagram for solving an incomplete system of linear equations with input parameter supplementation

After calculating the inverse transformation

$$\beta^{-1}: y \to [\beta_1^{-1}(y_1),...,\beta_l^{-1}(y_l)] = [x_1,...,x_l] \tag{7}$$

we obtain the information vector $x[l] = [x_1, x_2,...x_l]$.

Since the mapping $\sigma: y \to y^*$ leads to the calculation of $q$ vectors $y^* = [y_1^*,...,y_q^*]$, we should construct $q$ displays vector $x$ to vector $y^*$

$$\beta^*: x \to [\beta_1^*(x),...,\beta_q^*(x)] = [y_1^*,...,y_q^*], \tag{8}$$

where is the mapping $\beta_j^*(x) = y_j^*, j = \overline{1,q}$ defines a transformation of components for all vectors $x$

$$\beta_j^*: x \to [\beta_{j1}(x_1),...,\beta_{jk}(x_k)] = [y_{j1},...,y_{jk}] = y_j^*, \; j = \overline{1,q}. \tag{9}$$

Since the transformation $\sigma^{-1}$ in the secret sharing scheme is defined over a linear vector space, it can be transferred to the transformations $\beta_j^*$

$$\beta = \sigma^{-1}(\beta_1^*,...,\beta_q^*) \tag{10}$$

The block diagram for constructing mappings through homomorphic transformation is presented in Fig. 2.

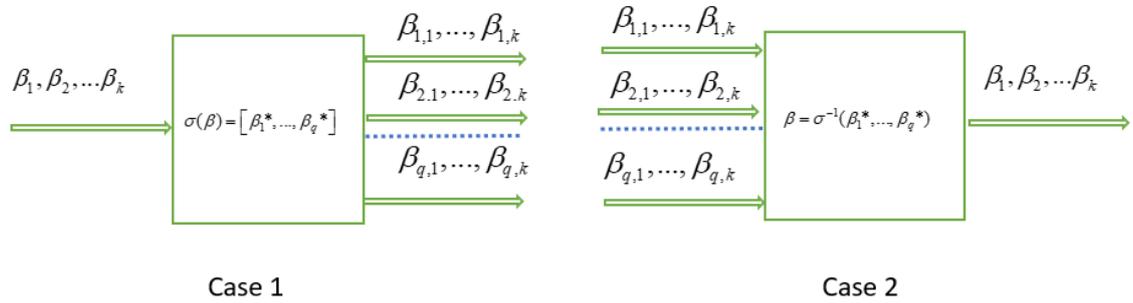

Figure 2 – Block diagram of homomorphic transformation $\sigma$ for mappings $\beta: x \to y$

Case 1 - action of direct transformation $\sigma$, Case 2 - action of inverse transformation $\sigma^{-1}$.

*The construction of a cryptosystem based on complete system of equations includes the following stages:*

Stage 1. Construction of transformation $\beta: x \to [y_1,...,y_k]$ (3)

Stage 2. Construction of a set of transformations for a homomorphic transformation $\sigma$: $\beta^*: x \to [y_1^*,...,y_q^*]$ (7) with the property that the inverse transformation action $\sigma^{-1}: y^* \to y$ for any input vectors leads to a fixed secret vector $y[k-l] = [y_{l+1}, y_{l+2},...y_k]$.

In a public key cryptosystem, the transformation $\beta_j^*: x \to y_j^*$ must be keyless. The shared secret $y[l] = [y_1, y_2,...y_l]$ is computed by solving equation (6). The inverse transformation $\beta^{-1}: y \to [x_1,...,x_l]$ (7) is bijective and secret. The transformation $\beta$ is $m$ bitwise substitutions for vector components $x$.

Secrecy of the LINE cryptosystem is achieved by the fact that it is possible to construct secret transformations $\sigma$, $\beta$ and the secret vector $y[l,k] = [y_{l+1}, y_{l+2},...y_k]$.

Lets consider the LINE cryptosystem parameters as follows.

1. We use the following **general parameters**: binary random matrix $A[l \times k]$, $A = A_1 \| A_2$, where $A_1[l \times l]$ is a non-singular matrix and $A_2[l \times (k-l)]$ is an arbitrary matrix, $\|$ concatenation of matrix rows.

2. We use the transformation vectors $\beta^* =: [\beta_1^*, ..., \beta_q^*]$ to create **public keys**.

3. We have transformations $\sigma$, $\beta$ and secret vector $y[l,k] = [y_{l+1}, y_{l+2}, ... y_k]$ as **secret keys**.

The secrecy of the cryptosystem is based on the secrecy of the secret sharing scheme. The implementation costs are determined by calculations using equations $(4) \div (7)$.

In the next section we will consider the construction of transformations $\beta$ based on secret one-way substitutions.

**Construction of one-way substitutions**

The $\beta$ and $\beta^*$ transformations in expressions (2) and (8) act as one-way functions on the elementary Abelian 2-group of order $2^m$. The requirement of asymmetry for public key encryption scheme determines that the direct transformations $\beta^*$ must be keyless and $\beta$ - secret.

The $\beta$ and $\beta^*$ transformations act as substitutions for $m$ bit strings. Three implementations of substitutions can be distinguished: tabular, analytic, and based on basis vectors. The tabular implementation requires $2^m$ words, which leads to the highest operational memory costs. Analytic substitution is calculated from expressions and is therefore not secret.

Let us consider the construction of substitutions with calculations based on basis vectors. The construction of transformations with such properties was introduced by Magliveras in his symmetric key cryptosystem PGM (Permutation Group Mappings) [14]. PGM cryptosystem built on group bases for finite permutation groups, which are known as logarithmic signatures. Later, Magliveras, Stinson, van Trung, Lempken and Wei proposed public- key cryptosystems based on group covers in MST1, MST2 and based on random coverings of finite non-Abelian groups in MST3 [15]. The ideas presented in MST3 were further developed for multiparametric groups [16,17]. All presented cryptosystems are based on group factorization of large finite groups. Encryption is performed based on encryption over group bases, while decryption is based on secret group factorization. Efficient group factorization directly affects the operational costs of cryptographic computations. As demonstrated by the results of designing the MST3 cryptosystem, key overhead reaches 1 Mbit and more, which reduces practical attractiveness [18].

We will construct transformations $\beta$ and $\beta^*$ as one-way permutations for an Abelian 2-group of order $2^m$. The group basis defines a vector space of dimension $m$ over $\beta^*$ the field $F_2$.

Let $\varsigma$ be elements of the Abelian group and be defined $m$ by bit strings. Let be $r = r_1, r_2, ..., r_m$ an input $m$ bit string. We define the bits $r_j$ of the string $r$ in the notation of spinors $\bar{r}_j = \begin{vmatrix} 1 - r_j \\ r_j \end{vmatrix}$. For bit $0$ we have a spinor $\bar{0} = \begin{vmatrix} 1 \\ 0 \end{vmatrix}$ and bit $1$ a spinor $\bar{1} = \begin{vmatrix} 0 \\ 1 \end{vmatrix}$.

We represent the factorization of an Abelian 2-group of order $2^m$ by a matrix $\beta$ of bit strings with pairwise blocks $\beta = [B_1, B_2, ..., B_m]$

$$\beta = \begin{vmatrix} B_1 \\ B_2 \\ \vdots \\ B_m \end{vmatrix} = \begin{vmatrix} b_{(11)_0}, b_{(12)_0}, ..., b_{(1m)_0} \\ b_{(11)_1}, b_{(12)_1}, ..., b_{(1m)_1} \\ b_{(21)_0}, b_{(22)_0}, ..., b_{(2m)_0} \\ b_{(21)_1}, b_{(22)_1}, ..., b_{(2m)_1} \\ \vdots \\ b_{(m1)_0}, b_{(m2)_0}, ..., b_{(mm)_0} \\ b_{(m1)_1}, b_{(m2)_1}, ..., b_{(mm)_1} \end{vmatrix} \qquad (11)$$

The calculation of the transformation $\beta$ for $m$ a bit word $r$ is conveniently defined by the tensor product

$$\beta(r) = |\bar{r}_1, \bar{r}_2, ..., \bar{r}_m| \otimes |B_1, B_2, ..., B_m| = \bar{r}_1 \otimes B_1 + \bar{r}_2 \otimes B_2 + ... + \bar{r}_m \otimes B_m, \qquad (12)$$

where

$$\bar{r}_j \otimes B_j = \begin{vmatrix} 1 - r_j \\ r_j \end{vmatrix} \otimes \begin{vmatrix} b_{(j1)_0}, b_{(j2)_0}, ..., b_{(jm)_0} \\ b_{(j1)_1}, b_{(j2)_1}, ..., b_{(jm)_1} \end{vmatrix} =$$

$$\left| b_{(j1)_0}(1 - r_j) + b_{(j1)_1} r_j \right|, \left| b_{(j2)_0}(1 - r_j) + b_{(j2)_1} r_j \right|, ..., \left| b_{(jm)_0}(1 - r_j) + b_{(jm)_1} r_j \right|$$

The block diagram for computing substitutions based on transformations $\beta$ is presented in Fig. 3.

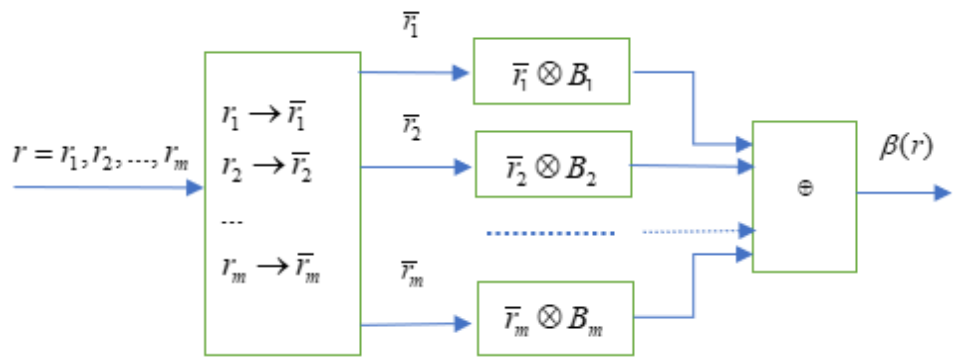

**Figure 3 - Scheme for computing transformation $\beta$ for $m$ - bit word $r$**

Let's consider an example of simple factorization. Let $m=4$ and be defined $\beta$ by the following matrix

$$\beta = \begin{vmatrix} B_1 \\ \hline B_2 \\ \hline B_2 \\ \hline B_m \end{vmatrix} = \begin{vmatrix} 0000 \\ 1000 \\ \hline 0000 \\ 0100 \\ \hline 0000 \\ 0010 \\ \hline 0000 \\ 0001 \end{vmatrix} \qquad (13)$$

For the string $r = 0110$ we calculate $z = \beta(r)$. We get a trivial result $z = 0110$

$$\beta(r) = \beta(0110) = |\bar{0}, \bar{1}, \bar{1}, \bar{0}| \otimes |B_1, B_2, B_3, B_4| = \bar{0} \otimes B_1 + \bar{1} \otimes B_2 + \bar{1} \otimes B_3 + \bar{0} \otimes B_4 =$$

$$\begin{vmatrix} 1 \\ 0 \end{vmatrix} \otimes \begin{vmatrix} 0000 \\ 1000 \end{vmatrix} + \begin{vmatrix} 0 \\ 1 \end{vmatrix} \otimes \begin{vmatrix} 0000 \\ 0100 \end{vmatrix} + \begin{vmatrix} 0 \\ 1 \end{vmatrix} \otimes \begin{vmatrix} 0000 \\ 0010 \end{vmatrix} + \begin{vmatrix} 1 \\ 0 \end{vmatrix} \otimes \begin{vmatrix} 0000 \\ 0001 \end{vmatrix} = |0110| \qquad (14)$$

Consider the inverse transformation $\beta^{-1} : z \to r$, $r = r_1, r_2, ..., r_m$. Let $z = 0110$ and be the matrix $\beta$ defined in (13). The most significant bit $b_4$ of the word is calculated by the rows of the block $B_4$. The value of the bit $b_4 = 0$ corresponds to the case when the first row was added to the sum (14) $B_4$. This determines the spinor $\bar{0}$ and $r_4 = 0$. From (14), we extract the component corresponding to the fourth spinor

$$z' = z + (\bar{0} \otimes B_4) = |0110|.$$

To determine the third bit, $r_3$ we apply the rows of the block $B_3$. For example, this will be the row $|0010|$ and the bit $r_3 = 1$. Extracting the component $|0010|$ corresponding to the third spinor from $z'$ gives

$$z'' = z' + (\bar{1} \otimes B_3) = |0100|.$$

We continue these actions iteratively until the last bit of the string is determined $r$.

Factoring a group by bases determines the structures and types of blocks of the matrix $\beta$. In the example considered, we used blocks of type 2, which determines two basis elements of the finite group in each block. This corresponds to a one-bit element of the row $x$. Blocks with a larger number of basic elements can be used. If the row $x$ is broken down into bit-by-bit elements $n$, then the basis blocks must be of type $2^n$, $n < m$. In this case, spinors of size should be used to calculate $\beta(x)$ in expression (12) $2^n$.

The direct transformation $\beta(x) = y$ is calculated using the tensor product of the input word and the matrix with the group bases. To calculate the inverse transformation, $\beta^{-1}(y) = x$ it is necessary to know the factorization $\beta$. The secrecy of the group factorization can be ensured by homomorphic transformations of the elements of the basic blocks, merging the basic blocks, their permutation, and permutation of the elements in the blocks. The efficiency of such transformations, operating costs, and the secrecy provided are widely discussed in [18].

We construct transformations $\beta$ based on a secret factorization over an Abelian 2-group of order $2^m$, using a set of secret homomorphic transformations [18]. Let $\beta_1 = [B_1, B_2, ..., B_m]$ is a prime factorization of an Abelian 2-group of order $2^m$ with blocks of type 2. The set of transformations of the group vectors for constructing the secret factorization is as follows:

- permutation of elements $\rho_1 : \beta_1 \to \beta_2$ in blocks $B_j$, $j = \overline{1, m}$;

- rearrangement $\rho_2 : \beta_2 \to \beta_3$ blocks in array $\beta_2$;

- adding random bits $\rho_3 : \beta_3 \to \beta_4$ to block rows $B_j$, $j = \overline{1, m}$;

- secret homomorphic transformation based on polynomial multiplication $\rho_4 : \beta_4 \to \beta_5$, $\beta_5 = \gamma \cdot \beta_4$ rows of blocks $B_j$, $j = \overline{1, m}$, where is $\gamma$ a polynomial $\gamma \in F(2^m)$;

- secret homomorphic transformation based on matrix multiplication $\rho_5 : \beta_5 \to \beta_6$, $\beta_6 = \beta_5 \cdot \psi$ rows of blocks $B_j$, $j = \overline{1, m}$, Where $\psi$ non-singular binary matrix of dimension $m \times m$.

As a result, we achieved the transformation $\beta = [B_1, B_2, ..., B_m]$.

**Let us consider an example.** Let us construct a factorization $\beta$ with blocks of bases of an Abelian 2-group of type 2. Let $m = 6$.

Let us define:

- a prime factorization of the group $\beta_1$, which is presented in Table 1;

- permutation matrix $\rho_1 =: [\ 110110]$ elements in the blocks of the matrix $\beta_1$;

- permutation matrix $\rho_2 =: [\ 340152]$ matrix $\beta_2$ blocks $[B_1, B_2, ..., B_m]$;

- random vectors $\upsilon = [\upsilon_1, \upsilon_2, ..., \upsilon_m]$, $\upsilon_j \in F(2^m)$, $j = \overline{1, m}$ to transform $\rho_3 : B_j(i)_4 = B_j(i)_3 + \upsilon_j$, $i = \overline{1, 2}$, $j = \overline{1, m}$

$$\upsilon = [\upsilon_1, \upsilon_2, ..., \upsilon_m] = \begin{vmatrix} 101111 \\ 101000 \\ 111001 \\ 010100 \\ 000000 \\ 011110 \end{vmatrix} ;$$

- random polynomial $\gamma = 1 + x + x^2 + x^4$ for $\rho_4 : B_j(i)_5 = B_j(i)_4 \cdot \gamma$, $i = \overline{1, 2}$, $j = \overline{1, m}$;

- non-degenerate bit matrix y $\psi$ for $\rho_5: B_j(i)_6 = B_j(i)_5 \cdot \psi$, $\quad i=\overline{1,2}$, $j=\overline{1,m}$

$$\psi_{m \times m} = \begin{vmatrix} 101000 \\ 001010 \\ 110001 \\ 000111 \\ 010000 \\ 111010 \end{vmatrix}$$

The transformations $\rho_3 \div \rho_5$ are defined by the following expressions

$$\rho_3: B_j(i)_4 = B_j(i)_3 + \upsilon_j,$$

$$\rho_4: B_j(i)_5 = B_j(i)_4 \cdot \gamma,$$

$$\rho_5: B_j(i)_6 = B_j(i)_5 \cdot \psi.$$

The results of the calculations $\beta$ by steps are presented in Table 1.

*Table 1 – Transformations $\rho_1 \div \rho_5$*

| $\beta = [B_1,...,B_m]$ | $\beta_1$ | $\beta_1 \to \beta_2$ | $\beta_2 \to \beta_3$ | $B_j(i)_3 + \upsilon_j$ | $B_j(i)_4 \cdot \gamma$ | $B_j(i)_5 \cdot \psi$ |
|---|---|---|---|---|---|---|
| $B_1$ | 000000<br>100000 | 100000<br>000000 | 000000<br>001000 | 101111<br>100111 | 001011<br>110101 | 011011<br>011111 |
| $B_2$ | 100000<br>010000 | 010000<br>100000 | 000100<br>110000 | 101100<br>011000 | 011011<br>100011 | 010001<br>000010 |
| $B_3$ | 000000<br>001000 | 000000<br>001000 | 111000<br>001001 | 000001<br>110000 | 101111<br>100111 | 110100<br>000101 |
| $B_4$ | 110000<br>000100 | 000100<br>110000 | 100000<br>000000 | 110100<br>010100 | 111000<br>000010 | 010011<br>010000 |
| $B_5$ | 100000<br>010110 | 010110<br>100000 | 010000<br>100000 | 010000<br>100000 | 011101<br>111010 | 000110<br>000011 |
| $B_6$ | 111000<br>001001 | 111000<br>001001 | 010110<br>100000 | 001000<br>111110 | 111110<br>111001 | 000100<br>101001 |

The computation of the transformation $\beta(r) = z$ for $m$ a bit word $r$ is determined by the tensor product (12). The transformations $\rho_3 \div \rho_5$ mask the factorization of the group.

The computation of the inverse transform $\beta^{-1}(z) = r$ is performed through inverse operations $\rho_3^{-1} \div \rho_5^{-1}$ with reduction to a row $z_3$ in a factorizable group

$$\beta_3: z_3 = z\psi^{-1}\gamma^{-1} + \upsilon_\Sigma,$$

where $\upsilon_\Sigma = \sum_{j=1}^{m} \upsilon_j$.

For a string $z_3$, we apply factorization by a simple group $\beta_1$

$$\beta^{-1}(z_3) = (r_1, r_2, ..., r_m)_3.$$

Let's get the original data string $r$ after inverse permutations $\rho_1, \rho_2$

$$\rho_1^{-1}\left(\rho_2^{-1}(r_1, r_2, ..., r_m)_3\right) = r_1, r_2, ..., r_m.$$

The scheme for computing the inverse transformation $\beta^{-1}(z) = r$ is presented in Fig. 4.

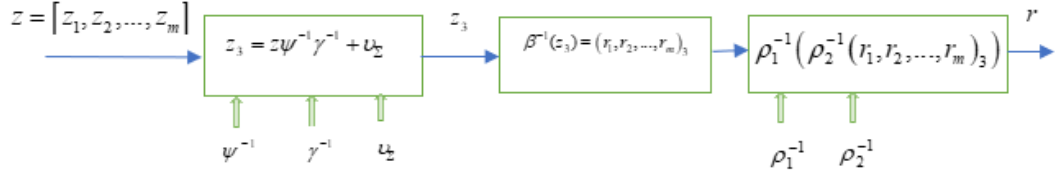

**Figure 4** - Scheme for computing the inverse transformation $\beta^{-1}$ for $m$-bit word $z$.

Substitutions at a length $m$ of bits have potentially good secrecy characteristics since their number has an estimate of $2^m!$.

The entropy estimate of the number of permutations based on group factorization is determined by randomizing transformations $\rho_1 \div \rho_5$ and is large even for small values $m$ [19].

For example, we can limit ourselves to the number of non-singular binary matrices $\psi$ in $\rho_5$, which has the estimate

$$N_5 = (2^m - 1)(2^m - 2)(2^m - 2^2) \cdots (2^m - 2^{m-1}) \approx 2^{m^2 - 2}.$$

The memory cost of group factorization-based substitutions is equal to $2m$ basis vectors, which is significantly less than that of table implementation-based substitutions.

Let us consider the construction of a secret sharing scheme in the LINE cryptosystem.

**Secret sharing in LINE cryptosystem**

We construct a secret sharing based on a homomorphic transformation that defines a set of transformations $[\beta_1^*,...,\beta_q^*]$ (10) with the property that the action of the inverse transformation $\sigma^{-1}: y^* \to y$ for any input vectors leads to a fixed secret vector $y[l,k] = [y_{l+1}, y_{l+2},...y_k]$. The direct transformation $\sigma$ acting on the substitution $\beta$ we write through the mapping

$$\sigma: \beta \to (\beta_1^*,...,\beta_q^*). \tag{15}$$

Display $\beta_j^*(x) = y_j^*$, $j = \overline{1,q}$ (9) acts as a transformation for all components of the vector $x$

$$\beta_j^*: x \to [\beta_{j1}(x_1),...,\beta_{jk}(x_k)] = [y_{j1},...,y_{jk}] = y_j^*, \ j = \overline{1,q}.$$

The substitutions $\beta_{ji}(x_i) = y_{ji}$, $j = \overline{1,q}$, $i = \overline{1,m}$ are in general not bijective. As an example, a secret transformation $\beta$ can be constructed using the following calculation

$$\beta = \sum_{j=1}^{q} \beta_j^* \omega_j, \tag{16}$$

where $\beta_j^* = [\beta_{j1},...,\beta_{jk}]$ are component wise permutations of the same type as $\beta$, $\omega_j$ are secret bit matrices of size $m \times m$. Matrix multiplications $\beta_j^* \omega_j$ are performed similarly to the transformation $\rho_5$ presented in Section 2.

Application of expression (16) leads to the following construction algorithm for $\beta_j^*$, $j = \overline{1,q}$.

*Algorithm of substitutions construction:*

1. We fix $k$ a component secret factorizable permutation $\beta = [\beta_1,...,\beta_k]$. To construct it, we use the mappings $\rho_1 \div \rho_5$ from Section 3.

2. We fix component wise permutations $\beta_j^* = [\beta_{j1},...,\beta_{jk}]$, $j = \overline{2,q}$ and let $\beta_{ji}$, $i = \overline{1,k}$ be random transformation matrices of the same type as $\beta$.

3. We fix the matrices $\omega_j$, $j = \overline{1,q}$ size $m \times m$ and let $\omega_1$ be the matrix of unity.

Let's calculate $\beta_1^* = \beta + \sum_{j=2}^{q} \beta_j * \omega_j$.

The block scheme of the substitutions construction algorithm is presented in Fig. 5.

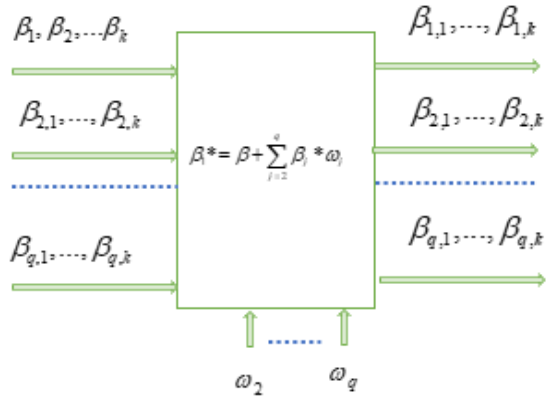

**Figure 5 – Algorithm for constructing substitutions $\sigma : \beta \to (\beta_1^*,...,\beta_q^*)$**

Encryption of $k$ the message component vector $x = [x_1, x_2, ... x_k]$ is determined by calculation

$$\beta_j^* : x \to [\beta_{j1}(x_1),...,\beta_{jk}(x_k)] = [y_{j1},...,y_{jk}] = y_j^*, j = \overline{1,q}.$$

Computing the shared secret

$$\sum_{j=1}^{q} y_j * \omega_j^{-1} = y.$$

Decryption is performed through the inverse transformation $\beta^{-1}(y) = x$.

Secret parameters: matrices $\omega_j$, $j = \overline{2,q}$.

The block diagram for computing private secrets and a shared secret with matrix secret transformation (16) is presented in Fig. 6.

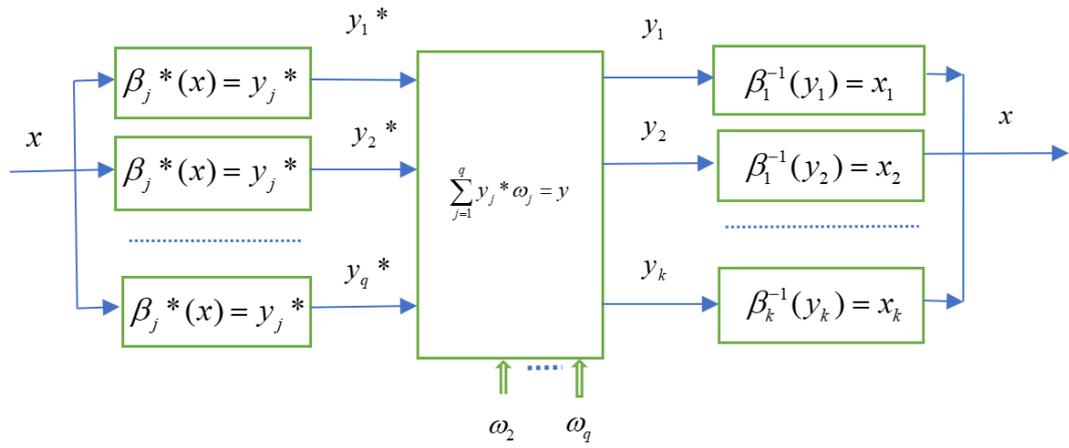

**Figure 6 - Algorithm for computing shared secret and decryption**

In the example considered, the shared secret is calculated from the set of matrices $\omega_j$, $j = \overline{2,q}$. For substitutions Even at small bit lengths, the secret sharing scheme (15) has potentially good privacy characteristics since matrix transformations of very high power can be constructed $|\omega_1||\omega_2|...|\omega_q|$. Let's build an public key encryption based on the LINE cryptosystem.

**LINE Public key encryption**

The implementation of public key encryption in the LINE cryptosystem is possible if the components $y[l,k] = [y_{l+1}, y_{l+2},...y_k]$ in the vector $y[k] = [y_1, y_2,...y_k]$ during decryption are determined for any input vector $x = [x_1, x_2,...x_k]$. Let $y[l,k]$ be the zero vector. Then, when calculating the homomorphic transformation, $\sigma$ we should obtain a vector $y[k]$ of the form

$$\sigma : \beta(x) = y[k] = [y_1, y_2,...y_l, 0,...,0]. \qquad (17)$$

To compute the secret homomorphic transformation $\sigma : \beta \to (\beta_1^*,...,\beta_q^*)$, where $\beta_j^* = [\beta_{j1},...,\beta_{jk}], j = \overline{1,q}$ we use the secret scheme (16). Then expression (17) will look like

$$\beta(x) = \sum_{j=1}^{q} \beta_j^*(x)\omega_j = [\beta_1^*(x), \beta_2^*(x),..., \beta_k^*(x)] = [y_1, y_2,...y_l, 0,...,0]. \qquad (18)$$

Expression (18) can be written taking into account the representation for each component $\beta_j^*, j = \overline{1,q}$ vectors $\beta$ in the form of

$$\beta_i(x) = \sum_{j=1}^{q} \beta_{ji}(x)\omega_j = y_i \quad i = \overline{1,k}.$$

To decrypt $[y_1, y_2,...y_l]$ substitutions $[\beta_1,...,\beta_l]$ must be factorizable. To construct factorizable permutations of type 2, we apply transformations $\rho_1 \div \rho_5$ over a basis of an Abelian 2 group of dimension $m$.

Let's define component wise substitutions $\beta_j^* = [\beta_{j1},...,\beta_{jk}]$, $j = \overline{2,q}$ as random matrices of the same type as permutations $[\beta_1,...,\beta_k]$.

We fix the matrices $\omega_j$, $j = \overline{1,q}$ size $m \times m$ and let $\omega_1$ be the matrix of unity.

Factorizable vector permutations $\beta$ - components $[\beta_1,...,\beta_l]$ are calculated via a homomorphic transformation of vectors $\beta_j^* = [\beta_{j1},...,\beta_{jl}]$, $j = \overline{1,q}$

$$\beta_i = \sum_{j=1}^{q} \beta_{ji} \omega_j , \; i = \overline{1,l} . \tag{19}$$

Given the condition that $\omega_1$ is the identity matrix, the vector $\beta_1^* = [\beta_{11},...,\beta_{1l}]$ is defined by the expression

$$\beta_1^* = \beta + \sum_{j=2}^{q} \beta_j * \omega_j ,$$

where $\beta = [\beta_1,...,\beta_l]$ factorizable permutations.

Let's define the components $[\beta_{l+1},...,\beta_k]$.

The mapping action $\sigma : \beta(x) \to y[k]$ for $[\beta_{l+1},...,\beta_k]$ results in a zero vector

$$[\beta_{l+1}(x),...,\beta_k(x)] = \left[ \sum_{j=1}^{q} \beta_{j(l+1)}(x)\omega_j, \sum_{j=1}^{q} \beta_{j(l+2)}(x)\omega_j,..., \sum_{j=1}^{q} \beta_{jk}(x)\omega_j \right] = [0,...,0]$$

substitute $i = \overline{l+1,k}$ the condition $\beta_i(x) = 0$ into expression (19) and obtain

$$\sum_{j=1}^{q} \beta_{ji} \omega_j = 0, \; i = \overline{l+1,k} . \tag{20}$$

The permutations $\beta_{ji}$, $j = \overline{2,q}$, $i = \overline{l+1,k}$ are defined as random matrices of the same type as the permutations $[\beta_1,...,\beta_l]$. From expression (20) we can calculate $i = \overline{l+1,k}$ the components of the vector $\beta_1^*$

$$\beta_1^* = \sum_{j=2}^{q} \beta_j * \omega_j .$$

Let's define $q$ secret vectors $\tau_j$ with matrix components $\tau_{ji}$, $i = \overline{1,k}$

$$\tau_j = [\tau_{j1},...,\tau_{jk}], \; j = \overline{1,q} .$$

Matrices $\tau_{ji}$ are bit-sized $[m \times m]$.

We define the sum of vectors $(\beta_j^* + \tau_j)$ as the component wise addition of matrices $\beta_{ji}$ and $\tau_{ji}$, $i = \overline{1,k}$. The matrix components $\beta_{ji}$ of the vector $\beta_j^*$ for type 2 contain $m$ blocks

$[B_1, B_2, ..., B_m]$ of two entries (11). The matrix components $\tau_{ji}$ of the vector $\tau_j$ contain $m$ entries $\tau_{ji} = \tau_{ji}[m]$. The sum of the matrices $\beta_{ji} + \tau_{ji}$ is determined by the bitwise addition of each row $\tau_{ji}[p]$ with entries in blocks $B_p$, $p = \overline{1, m}$

$$\beta_{ji} + \tau_{ji} = [B_{j1} + \tau_{j1}, B_{j2} + \tau_{j2}, ..., B_{jm} + \tau_{jm}], \qquad j = \overline{1,q} \qquad (21)$$

$$B_{js} + \tau_{js} = |B_{js}[1] + \tau_{js}[1], B_{js}[2] + \tau_{js}[2], ..., B_{js}[m] + \tau_{js}[m]|, \qquad s = \overline{1, m} \qquad (22)$$

$$B_{js}[p] + \tau_{js}[p] = \begin{vmatrix} b_{(p1)_0} + \tau_{p1}, b_{(p2)_0} + \tau_{p2}, ..., b_{(pm)_0} + \tau_{pm} \\ b_{(p1)_1} + \tau_{p1}, b_{(p2)_1} + \tau_{p2}, ..., b_{(pm)_1} + \tau_{pm} \end{vmatrix}_{js}, \qquad p = \overline{1, m} \qquad (23)$$

where $\tau_{p1}, \tau_{p2}, ..., \tau_{pm}$ are the bits of the string $\tau_{ji}[p]$.

Compute the transformations $(\beta_j * + \tau_j)$ for $j = \overline{1,q}$ each component of the input vector $x = [x_1, x_2, ...x_k]$ leads to the result

$$\beta_{ji}(x_i) + \tau_{ji}(x_i) = y_{ji} + \sum_{p=1}^{m} \tau_{ji}[p] = y_{ji} + \hat{\tau}_{ji}, \qquad 24)$$

where $\hat{\tau}_{ji}$ are m bit constants. Secret vectors $\tau_j$ with matrix components $\tau_{ji}$ are mapped to secret m bit vectors $\hat{\tau}_j = [\hat{\tau}_{j1}, ..., \hat{\tau}_{jk}]$ during calculation $\beta_j*(x) = y_j*$, $j = \overline{1,q}$.

Substitute (24) into expression (18) to calculate the shared secret

$$\beta(x) = \left[ \sum_{j=1}^{q}(\beta_{j1}(x) + \hat{\tau}_{j1})\omega_j, \sum_{j=1}^{q}(\beta_{j2}(x) + \hat{\tau}_{j2})\omega_j, ..., \sum_{j=1}^{q}(\beta_{jk}(x) + \hat{\tau}_{jk})\omega_j \right] =$$

$$\left[ \sum_{j=1}^{q}\beta_{j1}(x)\omega_j, \sum_{j=1}^{q}\beta_{j2}(x)\omega_j, ..., \sum_{j=1}^{q}\beta_{jk}(x)\omega_j \right] + \left[ \sum_{j=1}^{q}\hat{\tau}_{j1}\omega_j, \sum_{j=1}^{q}\hat{\tau}_{j2}\omega_j, ..., \sum_{j=1}^{q}\hat{\tau}_{jk}\omega_j \right] =$$

$$[\beta_1(x) + t_1, ..., \beta_k(x) + t_k] = [y_1 + t_1, y_2 + t_2, ...y_l + t_l, t_{l+1}, ..., t_k]$$

where $t_i = \sum_{j=1}^{q} \hat{\tau}_{ji}\omega_j$, $i = \overline{1,k}$ m bit components of the vector $t = [t_1, ..., t_k]$.

Vector $t = [t_1, ..., t_k]$ is a shared secret that is constructed from vectors $\hat{\tau}_j = [\hat{\tau}_{j1}, ..., \hat{\tau}_{jk}]$, $j = \overline{1,q}$.

Let's substitute (24) into expression (4) to calculate the cipher texts $[u_1, ..., u_q]$

$$A \times (y_1 * + \hat{\tau}_1) = u_1 + A \times \hat{\tau}_1 = u_1 + \hat{\tau}_{A1} = u_1'$$

$$A \times (y_2 * + \hat{\tau}_2) = u_2 + A \times \hat{\tau}_2 = u_2 + \hat{\tau}_{A2} = u_2'$$

$$..............$$

$$A \times (y_q * + \hat{\tau}_q) = u_q + A \times \hat{\tau}_q = u_q + \hat{\tau}_{Aq} = u_q'$$

Vectors $u_1', u_2', ..., u_l'$ consist of $l$ $m$ bit words.

The sequence of operations for encrypting the input vector can be represented by the block diagram in Fig. 7.

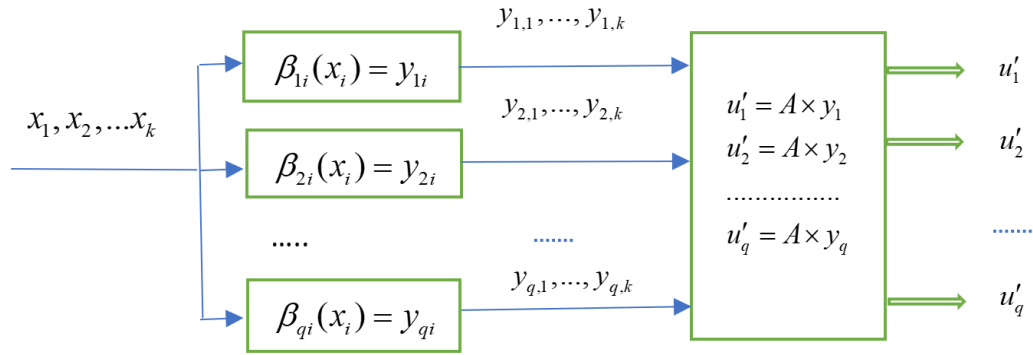

**Figure 7 - Encryption scheme of the LINE algorithm**

To decrypt, we calculate the transformation $\sigma$ for the same-named components of vectors $u_1', u_2', ..., u_l'$ and $\hat{\tau}_{A1}, \hat{\tau}_{A2}, ..., \hat{\tau}_{Aq}$

$$u_\sigma' = \sigma(u_1', u_2', ..., u_q'),$$

$$t_A = \sigma(\hat{\tau}_{A1}, \hat{\tau}_{A2}, ..., \hat{\tau}_{Aq}),$$

$$u_\sigma = u_\sigma' + t_A.$$

We get the equation for the cipher text

$$A \times y = A_1 \times y[l] = u_\sigma,$$

which has a solution for $y[l]$, since the components $y[l,k] = [y_{l+1}, y_{l+2}, ... y_k]$ are zero.

The block diagram of decryption is presented in Fig. 8.

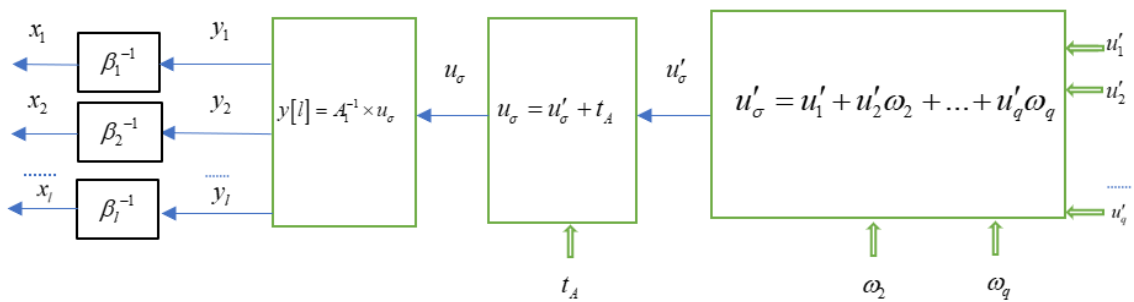

**Figure 8 - Decryption scheme of the LINE algorithm**

The actions described for encrypting the input vector $x = [x_1, x_2, ... x_k]$ and constructing a shared secret for decryption lead to the LINE public key encryption algorithm.

Let's consider the main steps of the algorithm. Here is the construction of general parameters.

1. Generate a binary random matrix $A[l \times k]$, $A = A_1 \| A_2$, where $l < k$, $A_1[l \times l]$ is a non-singular matrix and $A_2[l \times (k-l)]$ is an arbitrary matrix, $\|$ the concatenation of matrix rows.

2. Fix the parameters: $q$ - the number of substitutions, $m$ the number of bits for the components of the input vector $x$, the hash function $h$.

*We fix the following artifacts to construct the secret keys:*

- random binary matrices $\omega_j$, $j = \overline{1,q}$ dimensions $[m \times m]$ for homomorphic transformation $\sigma$;

- factorizable permutations $[\beta_1, ..., \beta_l]$ type 2 over Abelian 2 group of dimension $m$;

- random permutation vectors $\beta_j^* = [\beta_{j1}, ..., \beta_{jk}]$, $j = \overline{2,q}$ with matrix components $\beta_{ji}$, $i = \overline{1,k}$ type 2;

- random vectors of bit arrays $\tau_j = [\tau_{j1}, ..., \tau_{jk}]$, $j = \overline{1,q}$ dimensions $[m \times m]$, $\tau_{ji} = [\tau_{ji}[1], ..., \tau_{ji}[m]]$.

We calculate vectors as follows:

- $\hat{\tau}_j = [\hat{\tau}_{j1}, ..., \hat{\tau}_{jk}]$, $j = \overline{1,q}$, where $\hat{\tau}_{ji} = \sum_{p=1}^{m} \tau_{ji}[p]$, $i = \overline{1,k}$;

- $\hat{\tau}_A = [\hat{\tau}_{A1}, \hat{\tau}_{A2}, ..., \hat{\tau}_{Aq}]$, where $\hat{\tau}_{Aj} = A \times \hat{\tau}_j$, $j = \overline{1,q}$, are vectors $m$ of bit words of dimension $l$;

- $t_A = \sum_{j=1}^{q} \hat{\tau}_{Aj} \omega_j$ - vector of $m$ bit words of dimension $l$.

*We proceed with the next steps to construct the public keys:*

- $\beta'_{ji} = \beta_{ji} + \tau_{ji}$  $j = \overline{2,q}$,   $i = \overline{1,k}$; (25)

- $\beta'_{1i} = \beta_i + \sum_{j=2}^{q} \beta_{ji} \omega_j + \tau_{1i}$,   $i = \overline{1,l}$; (26)

- $\beta'_{1i} = \sum_{j=2}^{q} \beta_{ji} \omega_j + \tau_{1i}$,   $i = \overline{l+1,k}$. (27)

Addition with matrix components $\tau_{ji}$ is determined by expressions (21), (22), (23).

We consider the encryption stage with the following action. Let the message be defined $l$ by the components of the vector $x = [x_1, x_2, ... x_k]$ words and components $x_{l+1}, x_{l+2}, ... x_k$ are $m$ bit words from hashing $x_1, x_2, ... x_l$.

We compute:

- $y_j = [y_{j,1}, y_{j,2}, ..., y_{j,k}]$, $j = \overline{1,q}$, where $y_{j,i} = \beta'_{j,i}(x_i)$, $i = \overline{1,k}$, $j = \overline{1,q}$;

- $u'_j = A \times y_j$, $j = \overline{1,q}$.

We consider the decryption stage with the following action.

We compute

- $u'_\sigma = \sigma(u'_1, u'_2, ..., u'_q) = u'_1 + u'_2\omega_2 + u'_3\omega_3 + ... + u'_q\omega_q$;

- $u_\sigma = u'_\sigma + t_A$, $y[l] = A_1^{-1} \times u_\sigma$;

- $[\beta_1^{-1}(y_1), \beta_2^{-1}(y_2), ..., \beta_l^{-1}(y_l)] = [x_1, x_2, ..., x_l]$.

An example of computation of public key encryption is presented in the appendix.

**Secrecy of LINE public key encryption**

Let's analyze the secrecy of the LINE algorithm. First, let's consider brute force attacks.

**First attack.** An attack on a ciphertext with brute force against input messages has a complexity of $N_1 = 2^{(k-l)m}$.

The attack on the cipher text is determined by encrypting the input messages $x = [x_1, x_2, ...x_k]$ and comparing them with the known cipher text $u = [u_1, u_2, ..., u_q]$. The complexity of the attack will be determined by exhaustive search of the vector $x$ with complexity $2^{km}$. The result will be $2^{lm}$ equivalent solutions for $x$. The attack can be upgraded as follows. First, define the components of the vector $x = [x_{l+1}, x_{l+2}, ...x_k]$.

Then calculate the substitutions $\beta'_{j,i}(x_i) = y_{j,i}$, $i = \overline{l+1,k}$, $j = \overline{1,q}$, which gives the vectors $y_j = [y_{j,l+1}, y_{j,l+2}, ..., y_{j,k}]$, $j = \overline{1,q}$.

Let's compute the cipher texts

$u'_j = A_2 \times y_j$, $j = \overline{1,q}$.

Using decryption, we construct a vector $[x_{l+1}, x_{l+2}, ...x_k]$

$u''_j = u_j + u'_j$, $j = \overline{1,q}$,

$y[l] = A_1^{-1} \times u''$.

For components, $y[l]$ through inverse transformations, $[\beta_1^{-1}, \beta_2^{-1}, ..., \beta_l^{-1}]$ we calculate the components of the input vector

$[\beta_1^{-1}(y_1), \beta_2^{-1}(y_2), ..., \beta_l^{-1}(y_l)] = [x_1, x_2, ..., x_l]$.

Establishing a correspondence between the inputs and outputs of small substitutions is not a difficult task. We can assume that the last computation is feasible. The total number of searches can thus be reduced to $2^{(k-l)m}$.

An attack on a ciphertext is a brute-force problem with uncertainty equal to $(k-l)m$ bits in the solutions obtained. The success of the attack is determined by the probability of guessing $2^{-(k-l)m}$, which determines the secrecy of the cryptosystem at $s_T = (k-l)m$ the bit level.

**Second attack.** Attack on homomorphic transformation $\sigma(\beta_1,...,\beta_q,\omega)$ with matrix enumeration $\omega=[\omega_1,...,\omega_{q-1}]$, has complexity $N_2 = 2^{(q-1)m^2}$ and secrecy $s_\sigma = (q-1)m^2$, since it is determined by the number and dimension of secret matrices $\omega$.

The conditions for carrying out the attack are as follows. For two different input texts $x' = [x'_1, x'_2, ... x'_k]$ and $x'' = [x''_1, x''_2, ... x''_k]$ we calculate $\beta'_{ji}$, $j = \overline{1,q}$ for the components of the vectors $x'$ and $x''$

$$\beta'_{ji}(x'_i) = \beta_{ji}(x'_i) + \tau_{ji}(x'_i) = y'_{ji} + \sum_{p=1}^{m}\tau_{ji}[p] = y'_{ji} + \hat{\tau}_{ji}, \quad i = \overline{1,k} \qquad (28)$$

$$\beta'_{ji}(x''_i) = \beta_{ji}(x'') + \tau_{ji}(x'') = y''_{ji} + \sum_{p=1}^{m}\tau_{ji}[p] = y''_{ji} + \hat{\tau}_{ji}, \quad i = \overline{1,k}. \qquad (29)$$

We iterate over the matrices $\omega = [\omega_1,...,\omega_{q-1}]$ and calculate the common secret (22) by components $i = \overline{l+1,k}$ vectors (28) and (29) using formula (28)

$$\beta(x) = [y_1 + t_1, y_2 + t_2, ... y_l + t_l, t_{l+1}, ..., t_k]$$

where $t_i = \sum_{j=1}^{q}\hat{\tau}_{ji}\omega_j$, $i = \overline{1,k}$ $m$ bit components of the vector $t = [t_1,...,t_k]$. The search stops when the vectors $\beta(x')$ and $\beta(x'')$ coincide in components $[l+1,...,k]$.

**Third attack.** The attack on the key $t = [t_1,...,t_k]$ by enumerating $m$ bit vectors has complexity $N_3 = 2^{lm+l\log(2^m!)}$. Key vector attack $t = [t_1,...,t_k]$ is related to the secret homomorphism attack and aims to decipher the ciphertext to obtain the vector $y^* = [y_1, y_2, ..., y_l]$. Comparison with the known input message $x$ possible as a result of calculating secret inverse substitutions $\beta_i^{-1}(y_i) = x_i^*$, $i = \overline{1,l}$. The number of different substitutions $2^m!$ is very large even for small values of $m$. The secrecy of substitutions for an equiprobable choice for $l$ words of the input message vector is equal to $s = l\log(2^m!)$. If it is impossible to obtain a correspondence between the input and output words of a substitution, then the substitutions are secret. Conducting a brute-force attack on $t$ has complexity $2^{lm}$ and requires fixing the set of substitutions $\beta_1(x_1), \beta_2(x_2), ..., \beta_l(x_l)$. The total secrecy of the substitutions is equal to $s_\beta = lm + l\log(2^m!)$. Let's consider the analytical attacks.

**Fourth attack.** Analytical attack on $\omega = [\omega_1,...,\omega_q]$.

Let $q = 2$. The substitutions $\beta_{j,i}$ are defined by expressions (25), (26), (27) and have the following representation

$$\beta'_{1i} = \beta_i + \beta_{2i}\omega_2 + \tau_{1i}, \qquad i = \overline{1,l},$$

$$\beta'_{1i} = \beta_{2i}\omega_2 + \tau_{1i}, \qquad i = \overline{l+1,k},$$

$$\beta'_{2i} = \beta_{2i} + \tau_{2i}, \qquad i = \overline{1,k}.$$

Let us fix a vector $x = [x_1, x_2, \ldots x_k]$ and consider the values of the permutation vectors $\beta'_{1i}(x)$, $\beta'_{2i}(x)$. Let us write down the homomorphic transformation $\sigma$ for the components $x_i$, $i \in \overline{1,l}$

$$\sigma(x_i) := \beta'_{1,i}(x_i) + \beta'_{2,i}(x_i)\omega_2 = \beta_i(x_i) + \tau_{1,i}(x_i) + \tau_{2,i}(x_i)\omega_2 = \beta_i(x_i) + \hat{\tau}_{1,i} + \hat{\tau}_{2,i}\omega_2, \qquad i \in \overline{1,l}$$

The values $\tau_{1,i}(x_i)$, $\tau_{2,i}(x_i)$ are value invariant $x_i$ and are secret constants, as follows from (24)

$$\tau_{1,i}(x_i) = \hat{\tau}_{1,i},$$

$$\tau_{2,i}(x_i)\omega_2 = \hat{\tau}_{2,i}\omega_2.$$

Let's fix the components $x'_i$ and $x''_i$ calculate $\sigma(x'_i) + \sigma(x''_i)$

$$\sigma(x'_i) + \sigma(x''_i) := \beta'_{1,i}(x'_i) + \beta'_{2,i}(x'_i)\omega_2 + \beta'_{1,i}(x'') + \beta'_{2,i}(x'')\omega_2 = \beta_i(x'_i) + \beta_i(x''_i).$$

The expression for $\sigma(x'_i) + \sigma(x''_i)$ does not allow finding a solution with respect to $\omega_2$, since the value on the right-hand side $\beta_i(x'_i) + \beta_i(x''_i)$ is not known due to secrecy $\beta_i$, $i = \overline{1,l}$.

Let us consider a homomorphic transformation $\sigma$ for the components $x_i$, $i = \overline{l+1,k}$. Considering (9), we obtain

$$\sigma(x_i) := \beta'_{1,i}(x_i) + \beta'_{2,i}(x_i)\omega_2 = \hat{\tau}_{1,i} + \hat{\tau}_{2,i}\omega_2, \quad i \in \overline{l+1,k}$$

or equation

$$\beta'_{2,i}(x_i)\omega_2 + \hat{\tau}_{2,i}\omega_2 = \beta'_{1,i}(x_i) + \hat{\tau}_{1,i}.$$

For each column of $\omega_{2,n}$ the matrix $\omega_2$, $n = \overline{1,m}$ one can construct a system of linear equations for $k-l$ fixed values $x_i$, $i \in \overline{l+1,k}$

$$\begin{aligned}
\left(\beta'_{2,l+1}(x_{l+1}) + \hat{\tau}_{2,l+1}\right)\omega_{2,n} &= \left(\beta'_{1,l+1}(x_{l+1}) + \hat{\tau}_{1,l+1}\right)\omega_{1,n} \\
\left(\beta'_{2,l+2}(x_{l+2}) + \hat{\tau}_{2,l+2}\right)\omega_{2,n} &= \left(\beta'_{1,l+2}(x_{l+2}) + \hat{\tau}_{1,l+2}\right)\omega_{1,n} \\
&\ldots\ldots\ldots \\
\left(\beta'_{2,k}(x_k) + \hat{\tau}_{2,k}\right)\omega_{2,n} &= \left(\beta'_{1,k}(x_k) + \hat{\tau}_{1,k}\right)\omega_{1,n}
\end{aligned} \qquad (30)$$

where $n = \overline{1,m}$, $\omega_{1,n}$ is $n$ a column of the identity matrix $\omega_1$.

Let $k-l = m$ and the system of equations (30) have rank $m$. The system of equations (30) has a solution for fixed values of the vectors $\hat{\tau}_j = \left[\hat{\tau}_{j1}, \ldots, \hat{\tau}_{jk}\right]$, $j = \overline{1,2}$. The vectors $\hat{\tau}_j$, $j = \overline{1,2}$ are secret, then there is only a brute force attack with complexity $2^{2m^2}$. The attack is considered successful when for different components $x'_i$ and $x''_i$ the calculations on the left side of the equations (30) for the found matrix $\omega_2$ give the same values on the right parts.

Consider a cryptosystem with a homomorphic transformation $\sigma$ with two secret matrices $[\omega_2, \omega_3]$, case $q = 3$. The homomorphic transformation $\sigma$ for the components $x_i$ will $i \in \overline{1,l}$ have the following form

$$\sigma(x_i) =: \beta'_{1,i}(x_i) + \beta'_{2,i}(x_i)\omega_2 + \beta'_{3,i}(x_i)\omega_3 =$$
$$\beta_i(x_i) + \tau_{1,i}(x_i) + \tau_{2,i}(x_i)\omega_2 + \tau_{3,i}(x_i)\omega_3 = \beta_i(x_i) + \hat{\tau}_{1,i} + \hat{\tau}_{2,i}\omega_2 + \hat{\tau}_{3,i}\omega_3$$

where $i \in \overline{1,l}$.

The expression for $\sigma(x_i)$ does not allow finding a solution with respect to $\omega_2$ and $\omega_3$ since the value on the right side $\beta_i(x_i)$ is not known due to secrecy $\beta_i$, $i = \overline{1,l}$.

The homomorphic transformation $\sigma$ for the components has the $x_i$ following $i = \overline{l+1,k}$ representation

$$\sigma(x_i) =: \beta'_{1,i}(x_i) + \beta'_{2,i}(x_i)\omega_2 + \beta'_{3,i}(x_i)\omega_3 = \hat{\tau}_{1,i} + \hat{\tau}_{2,i}\omega_2 + \hat{\tau}_{3,i}\omega_3, i \in \overline{l+1,k}.$$

The following equation can be written

$$\left(\beta'_{2,i}(x_i) + \hat{\tau}_{2,i}\right)\omega_2 + \left(\beta'_{3,i}(x_i) + \hat{\tau}_{3,i}\right)\omega_3 = \beta'_{1,i}(x_i) + \hat{\tau}_{1,i}.$$

For columns $\omega_{2,n}$, $n = \overline{1,m}$ matrix $\omega_2$ and columns $\omega_{3,n}$, $n = \overline{1,m}$ matrix one $\omega_3$ can construct a system of linear equations for $k-l$ fixed values $x_i, i \in \overline{l+1,k}$

$$\left(\beta'_{2,l+1}(x_{l+1}) + \hat{\tau}_{2,l+1}\right)\omega_{2,n} + \left(\beta'_{3,l+1}(x_{l+1}) + \hat{\tau}_{3,l+1}\right)\omega_{3,n} = \left(\beta'_{1,l+1}(x_{l+1}) + \hat{\tau}_{1,l+1}\right)\omega_{1,n}$$
$$\left(\beta'_{2,l+2}(x_{l+2}) + \hat{\tau}_{2,l+2}\right)\omega_{2,n} + \left(\beta'_{3,l+2}(x_{l+2}) + \hat{\tau}_{3,l+2}\right)\omega_{3,n} = \left(\beta'_{1,l+2}(x_{l+2}) + \hat{\tau}_{1,l+2}\right)\omega_{1,n} \quad (31)$$
$$\ldots\ldots\ldots$$
$$\left(\beta'_{2,k}(x_k) + \hat{\tau}_{2,k}\right)\omega_{2,n} + \left(\beta'_{3,k}(x_k) + \hat{\tau}_{3,k}\right)\omega_{3,n} = \left(\beta'_{1,k}(x_k) + \hat{\tau}_{1,k}\right)\omega_{1,n}$$

where $n = \overline{1,m}$, $\omega_{1,n}$ is $n$ a column of the identity matrix $\omega_1$.

Let $k - l = m$, then the system of equations (31) has $m$ equations for $2m$ the bits of unknown columns $\omega_{2,n}$, $\omega_{3,n}$.

The system of equations (31) has a solution if it is supplemented $m$ with equations for other fixed values $x'_i$, $i \in \overline{l+1,k}$ and the values of the vectors, are fixed $\hat{\tau}_j = \left[\hat{\tau}_{j1}, ..., \hat{\tau}_{jk}\right]$. $j = \overline{1,3}$. The vectors $\hat{\tau}_j$, $j = \overline{1,3}$ are secret, then there is only a brute force attack with complexity $2^{3m^2}$. The attack is considered successful when for different components $x_i$, $x'_i$ and $x''_i$ the calculations on the left side of equations (41) for the found matrices $\omega_2$, $\omega_3$ give the same values on the right parts.

Analysis of the analytical attack shows that its complexity is $2^{qm^2}$ due to the secrecy of the vectors $\hat{\tau}_j = [\hat{\tau}_{j1}, ..., \hat{\tau}_{jk}]$, $j = \overline{1,q}$.

**Secrecy estimates and implementation costs**

Estimates of the implementation of a cryptosystem are determined by the costs of implementation, cipher texts, and performance. The implementation costs are determined by the costs of general parameters, public and private keys.

We consider general parameters of the cryptosystem:

- a binary random matrix $A[l \times k]$ that can be specified using a generator with z starting from an initial parameter in $n_A$ bits;

- homomorphic transformation parameter $q$, word size of bit permutation vectors $m$, hash function $h$.

Public keys are determined by the number and size of substitutions $\beta'_{ji} = \beta_{ji} + \tau_{ji}$, $j = \overline{2,q}$, $i = \overline{1,k}$. The permutations $[\beta_{j1}, ..., \beta_{jk}]$ are random matrices $j = \overline{2,q}$ $\beta_{ji}$, $i = \overline{1,k}$ type 2 over an Abelian 2 group of dimension $m$. Random vectors $\tau_j = [\tau_{j1}, ..., \tau_{jk}]$, $j = \overline{1,q}$ consist of bit arrays $\tau_{ji}$ of $i = \overline{1,k}$ dimension $[m \times m]$. Substitutions $\beta'_{ji}$, such as the sum $\beta_{ji}$ and $\tau_{ji}$ are also random and can be generated by the initial parameter in $n_p$ bits.

The substitutions $\beta'_{1i}$ are $i = \overline{1,k}$ constructed according to formulas (8), (9) and have a size of $n_\beta = 2km^2$ bits.

Secret keys are determined by the following artifacts:

- factorizable permutations $[\beta_1, ..., \beta_l]$ and have a size of $n_f = 2lm^2$ bits;

- matrices $\omega_j$, $j = \overline{1,q}$ dimensions $[m \times m]$ for homomorphic transformation $\sigma$ and have a size of $n_w = qm^2$ bits;

- is a vector $t_A$ of dimension $l$ and has a size of $n_t = lm$ bits.

*The costs of cipher texts* are defined as $u'_j$, $j = \overline{1,q}$ and have a size of $n_u = qlm$ bits.

*Secrecy scores* are determined by brute force attacks:

- attack on a ciphertext with $s_T = (k-l)m$ bit secrecy;

- attack on homomorphic transformation $\sigma(\beta_1, ..., \beta_q, \omega)$ with secrecy $s_\sigma = (q-1)m^2$;

- attack on the key vector $t_A$ of dimension $l$ with secrecy $s_t = lm$ and factorizable substitutions with secrecy $s_\beta = l\log(2^m!)$.

Table 2 presents cost estimates for common parameters and keys.

*Table 2 – Implementation costs for general parameters, public and secret keys*

| General parameters | | | Public keys: $\beta'_{ji}, \beta'_{1i}$ | | Secret keys: $\beta$, $\omega_j$, $t_A$ | | | |
|---|---|---|---|---|---|---|---|---|
| $m$ | $[l \times k]$ | $q$ | $n_A, n_p$ bit | $n_\beta = 2km^2$ bits / bytes | $n_f = 2lm^2$ bits / bytes | $n_w = qm^2$ bits / bytes | $n_t = lm$ bits / bytes | |
| 8 | 16x32 | 2 | 128 | 4096 / 512 | 2048 / 256 | 128 / 16 | 128 / 16 | |
| 8 | 16x32 | 3 | 128 | 4096 / 512 | 2048 / 256 | 192 / 24 | 128 / 16 | |
| 8 | 16x32 | 4 | 128 | 4096 / 512 | 2048 / 256 | 256/32 | 128 / 16 | |
| 8 | 32x48 | 2 | 128 | 6144/768 | 4096 / 512 | 128 / 16 | 256 / 32 | |
| 8 | 32x64 | 2 | 128 | 8192/1024 | 4096 / 512 | 128 / 16 | 256 / 32 | |
| 16 | 8 x 16 | 2 | 128 | 8192/ 1024 | 4096/512 | 512 / 64 | 128/16 | |
| 16 | 12 x 24 | 2 | 128 | 12288 / 1536 | 6144/768 | 512 / 64 | 192/24 | |
| 16 | 16x32 | 2 | 128 | 16384 / 2048 | 8192 / 1024 | 512 / 64 | 256 / 32 | |
| 16 | 16x32 | 3 | 128 | 16384 / 2048 | 8192 / 1024 | 768/96 | 256 / 32 | |
| 16 | 32 x 48 | 3 | 128 | 24576 /3072 | 16384 / 2048 | 768/96 | 512/64 | |
| 32 | 16x32 | 2 | 128 | 65536/8192 | 32768/4 0 96 | 2048/256 | 512/64 | |

Table 3 presents the costs of the cipher text, secrecy estimates, and the computation time for encryption and decryption.

*Table 3 – Implementation costs for ciphertext, secrecy and computations*

| General parameters of the cryptosystem | | | Test code size $u'_j$ bit/byte | Secrecy (bit) | | Computational time with reduction to one bit of cipher text or secrecy | | |
|---|---|---|---|---|---|---|---|---|
| | | | | Cipher text attack | Attack on homomorphic transformation | Encryption | Decryption using the substitution table | Decryption using factorizable substitution |
| $m$ | $[l \times k]$ | $q$ | $n_u = qlm$ | $s_T = (k-l)m$ | $s_\sigma = (q-1)m^2$ | sec/ $qlm$ | sec/ $\min(s_T, s_\sigma)$ | sec/ $\min(s_T, s_\sigma)$ |
| 8 | 16x32 | 2 | 256/32 | 128 | 64 | 1.01e-05 | 1.38e-06 | 3.58e-05 |
| 8 | 16x32 | 3 | 384/48 | 128 | 128 | 6.46e-06 | 7.32e-07 | 3.63e-05 |
| 8 | 32x64 | 2 | 512/64 | 256 | 64 | 1.80e-05 | 2.30e-06 | 7.46e-05 |
| 16 | 8x16 | 2 | 256/32 | 128 | 256 | 1.58e-06 | 3.28e-07 | 1.44e-05 |
| 16 | 12x24 | 2 | 384/48 | 192 | 256 | 2.38e-06 | 3.827e-07 | 1.76e-05 |
| 16 | 16x32 | 2 | 512/64 | 256 | 256 | 2.69e-06 | 5.22e-07 | 2.45e-05 |
| 16 | 16x32 | 3 | 768/96 | 256 | 512 | 1.69e-06 | 2.80e-07 | 2.52e-05 |
| 16 | 32x48 | 3 | 1536/192 | 512 | 512 | 2.32e-06 | 4.40e-07 | 4.54e-05 |

Estimates of the computational costs of executing the LINE algorithm with an implementation in Python using the NumPy library were performed on a MacBook Pro \ 2.0 GHz Dual - Core Intel Core i5.

The secrecy of the public ley encryption is determined primarily by the parameters of the incomplete system of linear equations and the homomorphic secret transformation on matrix calculations. Secret matrix transformations have potentially high entropy. We have considered a simple homomorphic transformation. It is possible to increase the security against attacks on the homomorphic transformation. It is possible to propose schemes based on multi-level constructions on matrix calculations and permutations. It is expected that the price for such solutions will be an increase in the cost of calculations and keys. Table 4 shows comparative characteristics of the costs of keys and cipher texts in bytes with known cryptosystems.

*Table 4 – Comparison analysis of LINE with present and PQC standards*

| Version | NIST Security | *SK* size | *PK* size | *CT* size |
|---|---|---|---|---|
| Kyber512 | AES128 | 1632 | 800 | 768 |
| Kyber768 | AES192 | 2400 | 1184 | 1088 |
| Kyber1024 | AES256 | 3168 | 1568 | 1568 |
| RSA3072 | AES128 | 384 | 384 | 384 |
| RSA15360 | AES256 | 1920 | 1920 | 1920 |
| LINE 128 ($m=8, k=32, l=16, q=3$) | AES128 | 288 | 528 | 48 |
| LINE 192 ($m=16, k=24, l=12, q=2$) | AES192 | 824 | 1536 | 48 |
| LINE 256 ($m=16, k=32, l=16, q=2$) | AES256 | 1088 | 2048 | 64 |

The key costs will be explained using the example of the LINE 128 cryptosystem with the parameters: m = 8, k = 32, l = 16, q = 3. The public key is determined by the costs of starting the random sequence generator to construct the matrix A and q - 1 random substitutions $\beta'_{ji} = \beta_{ji} + \tau_{ji}$ $j = \overline{2,q}$, $i = \overline{1,k}$ and the costs of transmitting 32 one-way substitutions $\beta'_{1i} = \beta_{1i} + \tau_{1i}$, $i = \overline{1,k}$. The costs of starting the generator are 16 bytes. The costs of transmitting 32 factorable one-way substitutions are 32x16x8 = 512 bytes, since each substitution is 16 single-byte records. Thus, the total costs will be 528 bytes. The cost of a secret key is determined by $q$ -1 secret matrices $\omega_j$, , $j = \overline{2,q}$ factorable permutation $[\beta_1,...,\beta_l]$ tables and a secret vector $t_A$ of dimension $l$. The cost of q -1 secret matrices $\omega_j$ is 16 bytes. The cost of a secret key $t_A$ is 16 bytes. The cost of one factorized substitution is 16 bytes, and given that there are 16 of them, we get 256 bytes. Thus, the total cost is 288 bytes. If we use only one factorized substitution, then the cost of secret keys will be 48 bytes.

**Conclusions**

Drawing from the substantial body of research in this domain [20-30], we introduce a novel and comprehensive approach to public key encryption. The LINE public key encryption cryptosystem based on solutions of linear equation systems with predefinition of input parameters through shared secret computation for factorizable substitutions is a good candidate for post-quantum cryptography. The application of an underdetermined system of linear equations for constructing public key encryption guarantees intractability with respect to input values. The distinction of LINE lies in the fact that no restrictions and conditions on data structures are imposed on the parameters, unlike other post-quantum cryptography candidates. The quantum security of LINE is based on high randomization of entries in arrays of factorized substitutions and the absence of any correlation in public parameters and ciphertexts. Through selection of cryptosystem public parameters, the declared NIST security levels of 128, 192, 256 bits and any other levels in general are achieved. The LINE algorithm scales well with respect to computational costs, memory, and hardware platform constraints without reducing the high level of security. The cost of public keys when computing over 8, 16, 32-bit words is in the range of 1-4 KB and is comparable with implementations for the best post-quantum cryptography candidates. Software implementation of the LINE algorithm for vectorized bitwise matrix computations can be very fast.

### Appendix 1 - An example of performing public key encryption

Let's consider an example for the following general parameters of the cryptosystem.

Let us define a system of linear equations by a matrix $A[l \times k]$ of dimension $l=6$, $k=12$.

$$A = A_1 \| A_2 = \begin{vmatrix} 110101111100 \\ 000011001010 \\ 010101111001 \\ 001101101111 \\ 110011101101 \\ 000111010010 \end{vmatrix}, \quad A_1 = \begin{vmatrix} 110101 \\ 000011 \\ 010101 \\ 001101 \\ 110011 \\ 000111 \end{vmatrix}, \quad A_2 = \begin{vmatrix} 111100 \\ 001010 \\ 111001 \\ 101111 \\ 101101 \\ 010010 \end{vmatrix}.$$

The matrix $A_1$ is non-singular and has an inverse matrix $A_1^{-1}$.

Let us define a cryptosystem for $q=2$ sets of ciphertexts $[u_1, u_2]$ With calculations over words $m=6$ bit and let $h$ is a hash function.

**Generating keys**

We generate

- random vectors of bit arrays $\tau_j = [\tau_{j,1},...,\tau_{j,12}]$, $j = \overline{1,2}$ dimensions $[6 \times 6]$. Please see the results of generation in Table A.1;

*Table A.1 – Generated random vectors of bit arrays $\tau_j = [\tau_{j,1},...,\tau_{j,12}]$*

| $\tau_{1,1} \div \tau_{1,12}$ | | | | | | | | | | | |
|---|---|---|---|---|---|---|---|---|---|---|---|
| 000111 | 100000 | 111100 | 010111 | 000010 | 111100 | 101100 | 110111 | 010011 | 100011 | 100011 | 100001 |
| 010101 | 101011 | 101100 | 101001 | 011011 | 110111 | 010000 | 101101 | 001011 | 011010 | 101000 | 000001 |
| 101100 | 011110 | 100000 | 010100 | 001101 | 010001 | 010011 | 000010 | 111110 | 000010 | 010000 | 110111 |
| 101001 | 010111 | 100001 | 100001 | 111001 | 001010 | 011101 | 111010 | 001011 | 100010 | 011000 | 010100 |
| 011010 | 011001 | 000100 | 100110 | 110000 | 010011 | 010010 | 110010 | 010101 | 010111 | 010001 | 011010 |
| 101011 | 010110 | 000001 | 000000 | 010111 | 100011 | 101100 | 001110 | 000011 | 001000 | 100010 | 011111 |
| $\tau_{2,1} \div \tau_{2,12}$ | | | | | | | | | | | |
| 001001 | 011110 | 101100 | 010011 | 111111 | 011111 | 110001 | 001110 | 001000 | 011001 | 001110 | 100000 |
| 010111 | 001110 | 101011 | 110001 | 001011 | 101001 | 100010 | 101010 | 010100 | 010101 | 001011 | 010011 |
| 110110 | 100110 | 010111 | 101011 | 001101 | 110000 | 100010 | 101000 | 110101 | 011000 | 101011 | 011001 |
| 110011 | 011011 | 110011 | 101000 | 101111 | 110011 | 110000 | 111111 | 011100 | 001010 | 010110 | 101100 |

| 010111 | 011111 | 110110 | 100101 | 010001 | 110101 | 010100 | 110111 | 100001 | 011001 | 100010 | 011011 |
| 100101 | 101111 | 010100 | 100010 | 100001 | 100010 | 110100 | 110010 | 111110 | 111011 | 111111 | 011100 |

- random binary matrix $\omega_2$ dimensions $[6 \times 6]$

$$\omega_2 = \begin{vmatrix} 010011 \\ 101110 \\ 010010 \\ 000011 \\ 010101 \\ 110001 \end{vmatrix}$$

We compute:

- $\hat{\tau}_j = [\hat{\tau}_{j1},...,\hat{\tau}_{jk}]$, $j = \overline{1,2}$, Where $\hat{\tau}_{ji} = \sum_{p=1}^{6} \tau_{ji}[p]$, $i = \overline{1,12}$;

- $\hat{\tau}_A = [\hat{\tau}_{A1}, \hat{\tau}_{A2}]$, where $\hat{\tau}_{Aj} = A \times \hat{\tau}_j$, $j = \overline{1,2}$, are vectors $m = 6$ of bit words of dimension $l = 6$;

- $t_A = \sum_{j=1}^{2} \hat{\tau}_{Aj} \omega_j$, a vector of $m = 6$ bit words of dimension $l = 6$. See Table A.2 for results.

Table A.2 – Computed results for $\hat{\tau}_j = [\hat{\tau}_{j1},...,\hat{\tau}_{jk}]$ and $\hat{\tau}_A = [\hat{\tau}_{A1}, \hat{\tau}_{A2}]$

| $\hat{\tau}_1$ | $\hat{\tau}_2$ | $\hat{\tau}_{A1}$ | $\hat{\tau}_{A2}$ | $\hat{\tau}_{A2}\omega_2$ | $t_A$ |
|---|---|---|---|---|---|
| 100110 | 101001 | 001001 | 110001 | 001100 | 000101 |
| 001101 | 011101 | 100001 | 001011 | 110110 | 010111 |
| 010100 | 000001 | 101111 | 100101 | 100001 | 001110 |
| 101101 | 100110 | 011110 | 010110 | 111000 | 100110 |
| 001010 | 100110 | 110110 | 000110 | 010110 | 100000 |
| 100000 | 100010 | 101001 | 110001 | 001100 | 100101 |
| 001100 | 100001 | | | | |
| 011110 | 110110 | | | | |
| 111011 | 101010 | | | | |
| 000110 | 111100 | | | | |
| 110000 | 100101 | | | | |
| 000110 | 000001 | | | | |

Let us construct permutations for an Abelian 2 group of dimension $m = 6$. Let all permutations be of type 2 and the same $\beta_i = \beta$. Let us take the factorizable permutation $\beta$ from the example in Section 3. Let's generate random permutation vectors $\beta_2^* = [\beta_{2,1},...,\beta_{2,12}]$, with matrix components of type 2. Results are shown in Table A.3.

Table A.3 – Generated random permutation vectors $\beta_2^* = [\beta_{2,1},...,\beta_{2,12}]$

| $\beta_{2,1} \div \beta_{2,12}$ | | | | | | | | | | | |
|---|---|---|---|---|---|---|---|---|---|---|---|
| 010011 | 000110 | 110101 | 100100 | 100101 | 111111 | 101110 | 111011 | 111101 | 001110 | 101111 | 000010 |
| 100011 | 010110 | 011110 | 110110 | 100000 | 110101 | 111101 | 001010 | 110110 | 001101 | 011101 | 100101 |
| 010010 | 110001 | 110100 | 001000 | 101101 | 101111 | 001100 | 001011 | 010000 | 111111 | 000111 | 110100 |
| 011111 | 100001 | 011010 | 011101 | 000001 | 101011 | 101000 | 001110 | 111100 | 010010 | 000001 | 010101 |
| 000111 | 100100 | 110000 | 100000 | 100011 | 100111 | 010000 | 001111 | 010000 | 010100 | 100101 | 001000 |
| 001110 | 001011 | 110111 | 000110 | 110101 | 010010 | 010011 | 100100 | 011011 | 100100 | 000011 | 111001 |
| 110100 | 010011 | 110010 | 010111 | 101010 | 100111 | 101111 | 001110 | 101001 | 010010 | 101101 | 001001 |
| 011101 | 111000 | 111000 | 111110 | 001001 | 110100 | 110001 | 101111 | 000000 | 100001 | 101010 | 111100 |
| 011011 | 111100 | 010110 | 100011 | 001010 | 010110 | 110101 | 000000 | 000110 | 110100 | 100011 | 011110 |
| 010011 | 000001 | 010100 | 001111 | 010000 | 100110 | 000110 | 101101 | 001100 | 000110 | 100001 | 101011 |
| 011001 | 000011 | 001100 | 110011 | 100011 | 100111 | 000110 | 101101 | 110000 | 110110 | 010010 | 010000 |
| 100010 | 111110 | 011100 | 011011 | 011010 | 011111 | 001111 | 011100 | 110010 | 111001 | 101111 | 101101 |

Let's compute $\beta'_{1i} = \beta + \beta_{2i}\omega_2 + \tau_{1i}$, $i = \overline{1,6}$ and $\beta'_{1i} = \beta_{2i}\omega_2 + \tau_{1i}$, $i = \overline{7,12}$. See Table A.4.

*Table A.4* – Computed $\beta'_{1i} = \beta + \beta_{2i}\omega_2 + \tau_{1i}$ and $\beta'_{1i} = \beta_{2i}\omega_2 + \tau_{1i}$

| $\beta'_{1,1} \div \beta'_{1,6}$ | | | | | | $\beta'_{1,7} \div \beta'_{1,12}$ | | | | | |
|---|---|---|---|---|---|---|---|---|---|---|---|
| 010110 | 101101 | 101000 | 011100 | 111000 | 101111 | 111011 | 111100 | 001110 | 100111 | 000101 | 110100 |
| 101111 | 000111 | 001001 | 100011 | 001110 | 101100 | 110001 | 110000 | 111000 | 000011 | 101101 | 000000 |
| 111111 | 110110 | 000011 | 101010 | 111001 | 000000 | 000001 | 011011 | 100101 | 010010 | 001111 | 111111 |
| 001100 | 001011 | 000111 | 100101 | 101000 | 010000 | 010001 | 101001 | 100111 | 100001 | 011001 | 011101 |
| 111111 | 111010 | 101001 | 110011 | 001110 | 010001 | 111101 | 110111 | 010000 | 101111 | 110001 | 100101 |
| 101101 | 101101 | 111111 | 000111 | 000111 | 101111 | 011001 | 010010 | 100110 | 010010 | 110100 | 101001 |
| 000100 | 001110 | 011010 | 111011 | 111110 | 101101 | 111011 | 111110 | 111011 | 011001 | 101011 | 110111 |
| 110111 | 101000 | 011110 | 001000 | 001010 | 100100 | 010001 | 011100 | 001011 | 000000 | 001100 | 111000 |
| 000100 | 110011 | 111010 | 010111 | 110001 | 101101 | 011101 | 110010 | 000011 | 101001 | 100110 | 110000 |
| 010011 | 101011 | 101010 | 010000 | 011101 | 010101 | 000100 | 000001 | 000100 | 000001 | 110011 | 111111 |
| 100010 | 110110 | 010100 | 011101 | 100100 | 010011 | 111010 | 111101 | 111110 | 100011 | 011001 | 110001 |
| 000100 | 000110 | 010111 | 110001 | 010111 | 010001 | 011001 | 110001 | 101011 | 010110 | 000100 | 101100 |

Then we compute $\beta'_{2i} = \beta_{2i} + \tau_{2i}$, $i = \overline{1,12}$. See Table A.5.

*Table A.5* – Computed $\beta'_{2i} = \beta_{2i} + \tau_{2i}$

| $\beta'_{2,1} \div \beta'_{2,12}$ | | | | | | | | | | | |
|---|---|---|---|---|---|---|---|---|---|---|---|
| 011010 | 011000 | 011001 | 110111 | 011010 | 100000 | 011111 | 110101 | 110101 | 010111 | 100001 | 100010 |
| 101010 | 001000 | 110010 | 100101 | 011111 | 101010 | 001100 | 000100 | 111110 | 010100 | 010011 | 000101 |
| 000101 | 111111 | 011111 | 111001 | 100110 | 000110 | 101110 | 100001 | 000100 | 101010 | 001100 | 100111 |
| 001000 | 101111 | 110001 | 101100 | 001010 | 000010 | 001010 | 100100 | 101000 | 000111 | 001010 | 000110 |
| 110001 | 000010 | 100111 | 001011 | 101110 | 010111 | 110010 | 100111 | 100101 | 001100 | 001110 | 010001 |
| 111000 | 101101 | 100000 | 101101 | 111000 | 100010 | 110001 | 001100 | 101110 | 111100 | 101000 | 100000 |
| 000111 | 001000 | 000001 | 111111 | 000101 | 010100 | 011111 | 110001 | 110101 | 011000 | 111011 | 100101 |
| 101110 | 100011 | 001011 | 010110 | 100110 | 000111 | 000001 | 010000 | 011100 | 101011 | 111100 | 010000 |
| 001100 | 100011 | 100000 | 000110 | 011011 | 100011 | 100001 | 110111 | 100111 | 101101 | 000001 | 000101 |
| 000100 | 011110 | 100010 | 101010 | 000001 | 010011 | 010010 | 011010 | 101101 | 011111 | 000011 | 110000 |
| 111100 | 101100 | 011000 | 010001 | 000010 | 000101 | 110010 | 011111 | 001110 | 001101 | 101101 | 001100 |
| 000111 | 010001 | 001000 | 111001 | 111011 | 111101 | 111011 | 101110 | 001100 | 000010 | 010000 | 110001 |

We obtain *public keys* : $\beta'_{j,i}$, $i = \overline{1,12}$, $j = \overline{1,2}$ and *secret keys* : $t_A$, $\omega_2$, $\beta$.

**Encryption**

Let the message be defined $l = 6$ by the components of the vector $x = [x_1, x_2, ... x_{12}]$ words and components $x_7, x_8, ... x_{12}$ are $m = 6$ bit words from hashing $x_1, x_2, ... x_6$.

Let's calculate: $y_j = [y_{j,1}, y_{j,2}, ..., y_{j,12}]$, $j = \overline{1,2}$, where $y_{j,i} = \beta'_{j,i}(x_i)$, $i = \overline{1,12}$, $j = \overline{1,2}$ and $u'_j = A \times y_j$, $j = \overline{1,2}$. Results are shown in Table A.6.

Table A.6 – Computed encryption data

| x | $y_1$ | $y_2$ | $u'_1$ | $u'_2$ |
|---|---|---|---|---|
| 111111 | 101110 | 110111 | 011100 | 011101 |
| 001111 | 110011 | 100110 | 010010 | 000100 |
| 111000 | 000101 | 011010 | 100001 | 110111 |
| 001101 | 011111 | 001010 | 001000 | 000111 |
| 001011 | 110010 | 111011 | 101101 | 010101 |
| 001101 | 011000 | 011101 | 011110 | 011001 |
| 010100 | 100001 | 110101 | | |
| 101010 | 111011 | 011101 | | |
| 101001 | 101000 | 001010 | | |
| 100100 | 110100 | 111001 | | |
| 001101 | 010000 | 101000 | | |
| 010001 | 100111 | 100100 | | |

**Decryption**

Let's make the following computations: $u'_\sigma = \sigma(u'_1, u'_2) = u'_1 + u'_2 \omega_2$, $u_\sigma = u'_\sigma + t_A$, $y[l] = A_1^{-1} \times u_\sigma$ and $[\beta_1^{-1}(y_1), \beta_2^{-1}(y_2), ..., \beta_l^{-1}(y_l)] = [x_1, x_2, ..., x_l]$. The results are presented in Table A.7.

Table A.7 – Computed decryption data

| $u'_2 \omega_2$ | $u'$ | $u_\sigma$ | $A_1^{-1}$ | $[y_1, y_2, ..., y_6]$ | $[x_1, x_2, ..., x_6]$ |
|---|---|---|---|---|---|
| 001110 | 010010 | 010111 | 101000 | 100010 | 111111 |
| 000011 | 010001 | 000110 | 111010 | 110101 | 001111 |
| 011010 | 111011 | 110101 | 110110 | 001001 | 111000 |
| 100111 | 101111 | 001001 | 010001 | 110000 | 001101 |
| 011100 | 110001 | 010001 | 110011 | 110110 | 001011 |
| 001101 | 010011 | 110110 | 100011 | 110000 | 001101 |

Let us check the calculations, for example for $\beta^{-1}(y_5^*) = x_5^*$. For this purpose, we calculate $\beta(x_5^*) = y_5^*$ using the substitution $\beta$ from the example in Section 3

$$\beta(x_5^*) = \beta(001011) =$$
$$|\bar{0},\bar{0},\bar{1},\bar{0},\bar{1},\bar{1}| \otimes |B_1, B_2, B_3, B_4, B_5, B_6| = \bar{0} \otimes B_1 + \bar{0} \otimes B_2 + \bar{1} \otimes B_3 + \bar{0} \otimes B_4 + \bar{1} \otimes B_5 + \bar{1} \otimes B_6 =$$
$$\begin{vmatrix}1\\0\end{vmatrix} \otimes \begin{vmatrix}011011\\011111\end{vmatrix} + \begin{vmatrix}1\\0\end{vmatrix} \otimes \begin{vmatrix}010001\\000010\end{vmatrix} + \begin{vmatrix}0\\1\end{vmatrix} \otimes \begin{vmatrix}110100\\000101\end{vmatrix} + \begin{vmatrix}1\\0\end{vmatrix} \otimes \begin{vmatrix}010011\\010000\end{vmatrix} + \begin{vmatrix}0\\1\end{vmatrix} \otimes \begin{vmatrix}000110\\000011\end{vmatrix} + \begin{vmatrix}0\\1\end{vmatrix} \otimes \begin{vmatrix}000100\\101001\end{vmatrix} = |110110|$$

Given the given parameters, the cryptosystem has the following cost estimates:

*Public keys* : $|\beta_{1,i}|$ - 864 $(2 \times m^2 \times k)$ bit,

128 bit to start the bit sequence generator to build $A$ and $\beta_{2,i}$.

*Secret keys* : $|t_A|$ - 36 $(l \times m)$ bit,

$|\omega_2|$ - 36 $(m \times m)$ bit,

$|\beta|$ - 72 $(2 \times m^2)$ bit.

Cipher text $\qquad [u'_1, u'_2]$ - 72 $(q \times l \times m)$ bat.

Secrecy $\qquad s$ - 36 $(m^2)$ bit.

Encryption time with reduction to one bit of the cipher text - 8.536709679497613 e -06.

Time to decrypt via table substitution with reduction to one bit of secrecy is 1.6623073154025608e-06.

Decryption time via factorized substitutions with reduction to one bit of secrecy is 1.728534698486328e -05.

## References

1. Alagic G., Apon D., Cooper D., Dang Q., Dang T., Kelsey J., Lichtinger J., Liu YK, Miller C., Moody D., Peralta R., Perlner R., Robinson A., Smith-Tone D.: Status Report on the Third Round of the NIST Post-Quantum Cryptography Standardization Process US Department of Commerce, NIST (2022)

2. Bos J, Ducas L, Kiltz E, Lepoint T, Lyubashevsky V, Schanck JM, Schwabe P, Seiler G, Stehle D (2018) CRYSTALS - Kyber: A CCA-secure module-lattice-based KEM. 2018 IEEE European Symposium on Security and Privacy (EuroS P), pp 353–367. https://doi.org/10.1109/EuroSP.2018.00032

3. Bindel N, Hamburg M, Hovelmanns K, H¨ulsing A, Persichetti E (2019) Tighter¨ proofs of CCA security in the quantum random oracle model. Theory of Cryptography, eds Hofheinz D, Rosen A (Springer International Publishing, Cham), pp 61–90.

4. Aguilar-Melchor C, Blazy O, Deneuville JC, Gaborit P, Zemor G (2018) Efficient en-´ cryption from random quasi-cyclic codes. IEEE Transactions on Information Theory 64(5):3927–3943. https://doi.org/10.1109/TIT.2018.2804444

5. Lyubashevsky V (2009) Fiat-Shamir with aborts: Applications to lattice and factoring-based signatures. Advances in Cryptology – ASIACRYPT 2009, ed Matsui M (Springer Berlin Heidelberg, Berlin, Heidelberg), pp 598–616.

6. Stehle D, Steinfeld R (2011) Making NTRU as secure as worst-case problems ´ over ideal lattices. Advances in Cryptology – EUROCRYPT 2011, ed Paterson KG (Springer Berlin Heidelberg, Berlin, Heidelberg), pp 27–47.

7. Ducas L, Lyubashevsky V, Prest T (2014) Efficient identity-based encryption over NTRU lattices. Advances in Cryptology – ASIACRYPT 2014, eds Sarkar P, Iwata T (Springer Berlin Heidelberg, Berlin, Heidelberg), pp 22–41.

8. Ducas L, Prest T (2016) Fast Fourier orthogonalization. Proceedings of the ACM on International Symposium on Symbolic and Algebraic Computation ISSAC '16 (Association for Computing Machinery, New York, NY, USA), p 191–198. https://doi.org/10.1145/2930889.2930923

9. Regev O (2005) On lattices, learning with errors, random linear codes, and cryptography. Proceedings of the Thirty-Seventh Annual ACM Symposium on Theory of Computing STOC '05 (Association for Computing Machinery, New York, NY, USA), p 84–93. https://doi.org/10.1145/1060590.1060603

10. Lyubashevsky V, Peikert C, Regev O (2010) On ideal lattices and learning with errors over rings. Advances in Cryptology – EUROCRYPT 2010, ed Gilbert H (Springer Berlin Heidelberg, Berlin, Heidelberg), pp 1–23.

11. Brakerski Z, Gentry C, Vaikuntanathan V (2012) (leveled) fully homomorphic encryption without bootstrapping. Proceedings of the 3rd Innovations in Theoretical Computer Science Conference ITCS '12 (Association for Computing Machinery, New York, NY, USA), p 309–325. https://doi.org/10.1145/2090236.2090262

12.] Banerjee A, Peikert C, Rosen A (2012) Pseudorandom functions and lattices. Advances in Cryptology – EUROCRYPT 2012, eds Pointcheval D, Johansson T (Springer Berlin Heidelberg, Berlin, Heidelberg), pp 719–737.

13. Kiltz E, Lyubashevsky V, Schaffner C (2018) A concrete treatment of Fiat-Shamir signatures in the quantum random-oracle model. Advances in Cryptology – EUROCRYPT 2018, eds Nielsen JB, Rijmen V (Springer International Publishing, Cham), pp 552–586.

14. Magliveras and N.D. Memon, "Algebraic properties of cryptosystem PGM", Journal of Cryptology, vol.5, no.3, pp.167–183, 1992.

15. W. Lempken, S. Magliveras, Tran van Trung and W. Wei, "A public key cryptosystem based on non- abelian finite groups", J. of Cryptology, 22 (2009), 62–74.

16. Khalimov, G., Kotukh, Y., Chang, S.-Y., Balytskyi, Y. Khalimova, S., Marukhnenko, O. "Encryption Scheme Based on the Generalized Suzuki 2-groups and Homomorphic Encryption " Communications in Computer and Information Science, 2022, 1536 CCIS, P. 59–76.

17. Khalimov, G., Kotukh, Y., Didmanidze, I., Khalimova, S., Vlasov, A. "Towards three-parameter group encryption scheme for MST3 cryptosystem improvement", Proceedings of the 2021 5th World Conference on Smart Trends in Systems Security and Sustainability, WorldS4 2021, 2021, страницы 204–211.

18. P. Svaba , "Covers and logarithmic signatures of finite groups in cryptography", Dissertation, https://bit.ly/2Ws2D24

19. Khalimov, G., Kotukh, Y., Kolisnyk, M., ... Sievierinov, O., Korobchynskyi, M. "Digital Signature Scheme Based on Linear Equations", Lecture Notes in Networks and Systems, 2025, 1285 LNNS, P. 711–728, 2025 Future of Information and Communication Conference, FICC 2025 Berlin 28 - 29 April 2025.

20. G. Khalimov, Y. Kotukh, Yu. Serhiychuk, O. Marukhnenko. "Analysis of the implementation complexity of cryptosystem based on Suzuki group" Journal of Telecommunications and Radio Engineering, Volume 78, Issue 5, 2019, pp. 419-427. DOI: 10.1615/TelecomRadEng.v78.i5.40

21. Y. Kotukh. "On universal hashing application to implementation of the schemes provably resistant authentication in the telecommunication systems" Journal of Telecommunications and Radio Engineering, Volume 75, Issue 7, 2016, pp. 595-605. DOI: 10.1615/TelecomRadEng.v75.i7.30

22. Kotukh, Y., & Khalimov, G. Hard Problems for Non-abelian Group Cryptography, 2021. In *Fifth International Scientific and Technical Conference" Computer and Information systems and technologies". https://doi.org/10.30837/csitic52021232176*.

23. Kotukh, Y., Severinov, E., Vlasov, O., Tenytska, A., & Zarudna, E. (2021). Some results of development of cryptographic transformations schemes using non-abelian groups. *Radiotekhnika*, *1*(204), 66-72.

24. Khalimov, G. *et al.* (2022). Encryption Scheme Based on the Generalized Suzuki 2-groups and Homomorphic Encryption. In: Chang, SY., Bathen, L., Di Troia, F., Austin, T.H., Nelson, A.J. (eds) Silicon Valley Cybersecurity Conference. SVCC 2021. Communications in Computer and Information Science, vol 1536. Springer, Cham. https://doi.org/10.1007/978-3-030-96057-5_5

25. G. Khalimov, O. Sievierinov, S. Khalimova, Y. Kotukh, S. -Y. Chang and Y. Balytskyi, "Encryption Based on the Group of the Hermitian Function Field and Homomorphic Encryption," *2021 IEEE 8th International Conference on Problems of Infocommunications, Science and Technology (PIC S&T)*, Kharkiv, Ukraine, 2021, pp. 465-469, https://doi.org/10.1109/PICST54195.2021.9772219 .

26. Gennady Khalimov, Yevgen Kotukh, Ibraim Didmanidze, and Svitlana Khalimova. 2021. Encryption scheme based on small Ree groups. In Proceedings of the 2021 7th International Conference on Computer Technology Applications (ICCTA '21). Association for Computing Machinery, New York, NY, USA, 33–37. https://doi.org/10.1145/3477911.3477917

27. Котух, Є. В. (2021). Кібербезпека у публічному секторі: монографія. *Харків: Колегіум*, *271*, 23-10.

28. Kotukh, E. ., Severinov, O. ., Vlasov, A. ., Kozina, L. ., Tenytska, A. ., & Zarudna , E. . (2021). Methods of construction and properties of logariphmic signatures . *Radiotekhnika*, *2*(205), 94–99. https://doi.org/10.30837/rt.2021.2.205.09

29. Kotukh, Ye. Method of Security Improvement for MST3 Cryptosystem Based on Automorphism Group of Ree Function Field / Yevgen Kotukh, Gennady Khalimov, Maxim Korobchinskiy // Theoretical and Applied Cybersecurity : scientific journal. – 2023. – Vol. 5, Iss. 2. – Pp. 31–39.

30. Kotukh, Y., Khalimov, G. Towards practical cryptoanalysis of systems based on word problems and logarithmic signatures. Retrieved from https://www.au.edu.az/userfiles/uploads/5231c8030469fa9a4b03963911a330d9.pdf

31. Khalimov, G., Kotukh, Y., Sergiychuk, Y., & Marukhnenko, A. (2018). Analysis of the implementation complexity of the cryptosystem on the Suzuki group. Radiotekhnika, 2(193), 7581. https://doi.org/10.30837/rt.2018.2.193.08

32. Kotukh, Y. ., Okhrimenko, T. ., Dyachenko, O. ., Rotaneva, N. ., Kozina, L. ., & Zelenskyi, D. . (2021). Cryptanalysis of the system based on word problems using logarithmic signatures. Radiotekhnika, 3(206), 106–114. https://doi.org/10.30837/rt.2021.3.206.09